\documentclass{jfm}

\usepackage{graphicx}
\usepackage{graphicx}
\usepackage{tikz}
\usepackage{epstopdf, epsfig}
\usepackage{newtxtext}
\usepackage{newtxmath}
\usepackage{natbib}
\usepackage{hyperref}
\hypersetup{
    colorlinks = true,
    urlcolor   = blue,
    citecolor  = black,
}

\newcommand{\RomanNumeralCaps}[1]
\linenumbers

\title{A finite-size correction model for two-fluid large-eddy simulation of particle-laden boundary layer flow}

\author{Antoine Mathieu\aff{1}\corresp{\email{antoine.mathieu@univ-grenoble-alpes.fr}},
  Julien Chauchat\aff{1},
  Cyrille Bonamy\aff{1},
  Guillaume Balarac\aff{1}\aff{,2}
 \and Tian-Jian Hsu\aff{3}}

\affiliation{\aff{1}LEGI, University of Grenoble Alpes, G-INP, CNRS, 38000 Grenoble, France
\aff{2}Institut Universitaire de France (IUF), 75005 Paris, France
\aff{3}Center for Applied Coastal Research, University of Delaware, Newark, DE 19716, USA}

\begin{document}
\maketitle

\begin{abstract}
In this paper the capabilities of the turbulence-resolving Eulerian–Eulerian two-phase flow model to predict the suspension of mono-dispersed finite-sized solid particles in a boundary layer flow are investigated. For heavier-than-fluid particles, having settling velocity of the order of the bed friction velocity, the two-fluid model significantly under-estimates the turbulent dispersion of particles. It is hypothesized that finite-size effects are important and a correction model for the drag law is proposed. This model is based on the assumption that the turbulent flow scales larger than the particle diameter will contribute to the resolved relative velocity between the two phases, whereas eddies smaller than the particle diameter will have two effects: (i) they will reduce the particle response time by adding a sub-particle scale eddy viscosity to the drag coefficient, and (ii) they will contribute to increase the production of granular temperature. Integrating finite-size effects allows us to quantitatively predict the concentration profile for heavier-than-fluid particles without any tuning parameter. The proposed modification of the two-fluid model extends its range of applicability to tackle particles having a size belonging to the inertial range of turbulence and allows us to envision more complex applications in terms of flow forcing conditions, \textit{i.e.} sheet flow, wave-driven transport, turbidity currents and/or flow geometries, \textit{i.e.} ripples, dunes, scour.
\end{abstract}


\section{Introduction}

Dispersed two-phase flows are present in many industrial and geophysical applications such as fluidized beds, slurry flows or sediment transport. Our ability to predict the dynamics of the system as a whole relies on our understanding of the fine-scale physical processes such as particle–particle interactions or fluid–particle interactions. One of the key challenges is the coupling between the particles and the carrier phase turbulence, the so-called turbulence-particle interactions \citep{balachandar2010}. The modelling methodology has to be carefully chosen depending on the available computational resources, flow regime and turbulence-particle interaction regime. 

For particles having a response time $t_s$ smaller than the Kolmogorov time scale $t_\eta$ associated with the smallest turbulent scales $\eta$ (typically $St<0.2$ with $St=t_s/t_\eta$ the Stokes number), the particles will follow almost exactly the carrier phase turbulence at all scales. For this regime, the equilibrium-Eulerian approach is a good approximation to model the particles dynamics and only mass and momentum conservation equations for the carrier phase are solved together with a relaxation equation for the particle phase velocity and the particle phase mass conservation equation \citep{ferry2005}. In many geophysical or industrial flows, the Stokes number may exceed 0.2 ($St>0.2$) and the particles no longer follow the carrier phase turbulence exactly \citep{balachandar2010}. In this situation, more sophisticated methodologies such as fully resolved direct numerical simulation (DNS), Eulerian-Lagrangian point–particle models or Eulerian–Eulerian two-fluid models are required to take into account the couplings between the particles and the carrier phase turbulence (two-way coupling) and the particle-particle interactions (four-way coupling). 

The most accurate method to account for turbulence-particle interactions is the fully resolved DNS \citep[e.g.][]{kidanemariam2013,vowinckel2014,vowinckel2017}. The interface between the carrier phase and the particles and, by extension, the fluid–particle interactions are explicitly resolved. In order to use this method, two constraints on the grid need to be satisfied: (i) the grid size needs to be everywhere of the order of the Kolmogorov length scale $\eta$ and (ii) the grid size should not be larger than one tenth of the particle size $d_p$ ($\Delta\sim d_p/10$ with $d_p$ the particle diameter). Putting together these constraints, fully resolved particle-laden boundary layer flow DNS is only achievable for bulk Reynolds numbers up to approximately $10^3$, with at most a few million particles and a billion grid points. In order to achieve Reynolds numbers relevant to realistic sediment transport conditions $O(10^5)$ for medium to very coarse sand, simulations would require of the order of $10^{12}$ to $10^{14}$ grid points. Such simulations are not possible with today's computational resources, and a compromise has to be found in terms of modelling strategy.

Concerning the Lagrangian point-particle approach, particles are considered punctual, their interactions with the carrier phase are modelled and Newton's second law is used to predict their trajectories \citep{maxey1983}. The limitations are twofold, on the one hand, the computational grid size $\Delta$ has to be much greater than the particle size and, on the other hand, the domain size is limited by the maximum number of particles achievable in the simulation. As an example, for the two-fluid simulation of scour around cylinders at the laboratory scale by \cite{mathieu2019} and \cite{nagel2020}, the number of particles involved would be on the order of 2 billion which is beyond the current computational power capacity. Furthermore, a separation of scale has to be satisfied and particles should be smaller than the Kolmogorov length scale ($d_p/\eta<1$). For finite-size particles ($d_p/\eta >1$), sub-particle scale processes need to be modeled in order to accurately predict the particle dynamics \citep{finn2016}.

Contrary to the fully resolved DNS and the Lagrangian point-particle methodology, the Eulerian–Eulerian two-fluid model has no limitations in term of number of particles. According to \cite{finn2016}, for particle-laden boundary layer simulations, the two-fluid approach is only suited for a narrow range of Stokes numbers $0.2 < St < 1$. Indeed for $St > 1$, the uniqueness of the Eulerian particle phase velocity field is not guaranteed \citep{ferry2001}. In other words, for a given fluid phase velocity field, the particles can follow different paths (\textit{i.e.} the particles velocity field is not unique) depending on the initial condition. Nevertheless, uniqueness of the particle phase velocity is not crucial considering time-averaged particle phase quantities (\textit{e.g.} concentration, velocity) and assuming ergodicity. More precisely, time-averaged variables are issued from multiple realization of the flow and, therefore, multiple particles trajectories. However, similarly to the point-particle approach, for finite-sized particles, additional sub-particle scale correction models are required \citep{finn2016}.

Over the last three decades, turbulence-resolving two-fluid models have been developed to simulate fluidized beds \citep{obrien1993, agrawal2001, heynderickx2004, wang2009, igci2008, ozel2013}. In this context, the particles are usually inertial ($St>1$) and smaller than the Kolmogorov length scale ($d_p/\eta<1$). The clear separation of scale between the fluid flow and the particles allows us to perform two-fluid DNS to fully resolve the turbulent spectrum without approximation. In fluidized beds particles show preferential concentration behaviour resulting in the formation of mesoscale structures such as clusters or streamers that can be captured by the two-fluid model \citep{agrawal2001}. Such structures have length scales of the order of 10 to 100 particle diameters and significantly impact the flow dynamics at large scale \citep{agrawal2001}. When performing large-eddy simulation (LES) in the framework of the two-fluid model, the effect of the unresolved mesoscale structures needs to be incorporated through sub-grid scale closures to accurately predict the two-phase flow dynamics \citep{agrawal2001}. Several sub-grid models have been tested by \cite{ozel2013} in this context and the functional model for the sub-grid drag force has been shown to perform better. Furthermore, \cite{ozel2013} showed that the effect of unresolved mesoscale structures vanishes for filter width $\Delta$ of the order of the particle diameter.

Recently, \cite{cheng2018} applied the two-fluid LES approach with the functional sub-grid drag model from \cite{ozel2013} to reproduce the unidirectional sheet flow experiment from \cite{revil2015}. The major difference between fluidized bed configurations mentioned above and the sheet flow configuration comes from the fact that, in the latter, particles are finite sized ($d_p/\eta>1$). In order to obtain accurate predictions of the flow and the particles dynamics, \cite{cheng2018} had to use a grid size slightly smaller than the particle diameter ($d_p/\Delta\geq1$). This simulation allowed us to explain, among other things, the physical origin of the modulation of the carrier phase turbulence induced by the presence of particles as being due to the turbulent drag work. However, \cite{cheng2018} observed an under-estimation of the time averaged sediment concentration in suspension and a strong sensitivity of the simulation results to the grid resolution. The sub-grid drag model from \cite{ozel2013} was originally designed to take into account the effect of unresolved particle clusters and streamers of the order of 10 to 100 particle diameters for coarse-grid simulations ($d_p/\Delta <0.1$) and not the effect of mesoscale structures for over-resolved simulations ($d_p/\Delta\geq1$). Therefore, the sub-grid closure used by \cite{cheng2018} was probably not ideal for this situation, thus explaining the under-prediction of the sediment concentration and the strong sensitivity to the grid resolution. As mentioned earlier, particles are bigger than the Kolmogorov length scale in this configuration. Finite-size effects probably play an important role and shall be modeled.

The modelling of interactions between the carrier phase and finite-size particles has been extensively studied in the literature. Experimental studies \citep{voth2002, qureshi2007, xu2008} provided evidence that finite-size particle dynamics is substantially affected by turbulent flow scales smaller than the particles compared with particles smaller than the Kolmogorov length scale. All the studies agreed on the facts that the variance of the acceleration probability density functions decreases for increasing particles size. These experimental observations have been further confirmed by numerical studies \citep{voth2002, calzavarini2009, homann2010, gorokhovski2018}. One way to recover some of the features of experimental and numerical finite-size particle acceleration probability density functions with the point–particle methodology is to include the Fax\'en correction term in the fluid-particle interaction model \citep{calzavarini2009}. The Fax\'en correction term takes into account the non-uniformity of the flow at the particle scale. While this method is suitable for Lagrangian simulations, the methodology developed by \cite{gorokhovski2018}, taking into account finite-size effect through an effective viscosity at the particle scale included in the expression of the drag force, is more suitable for volume-averaged two-phase flow models. 


In this paper, the two-fluid LES approach is applied to dilute suspension of finite-sized particles transported in a turbulent boundary layer flow. A finite-size correction model for the two-fluid approach inspired from the model proposed by \cite{gorokhovski2018} will be developed and tested against experimental data for particle-laden boundary layer flow configurations having $d_p/\eta >1$. In \S2, the two-fluid LES model formulation is presented. In \S3, the numerical results for one clear water configuration and three particle-laden flow configurations are presented with and without the finite-size correction model. In \S4, the sensitivity of the model results to the finite-size correction model components, grid resolution and second filter size are discussed. Finally, conclusions are drawn in \S5.

\section{Model formulation}
\label{tpmodel_eq}
\subsection{Filtered two-phase flow equations}

To perform LES with a two-phase flow model, a separation between the large turbulent flow scales (low frequency) and the small ones (high frequency) is operated by a filter (operator $\bar \cdot$). In analogy with compressible flows, a change of variable called Favre filtering is used to obtain filtered two-phase flow equations. Any variable $\psi(x_i, t)$ with $x_i=(x, y, z)^T$ the position vector and $i=1, 2, 3$ representing three spatial components can be decomposed into the sum $\psi(x_i, t) = \tilde\psi(x_i, t) + \psi^{''}(x_i, t)$, with $\tilde\psi(x_i, t)$ the resolved Favre-filtered part and $\psi^{''}(x_i, t)$ the unresolved sub-grid part. Favre-filtered fluid and solid velocities, $\tilde u^f_i = (\tilde u^f, \tilde v^f, \tilde w^f)^T$ and $\tilde u^s_i = (\tilde u^s, \tilde v^s, \tilde w^s)^T$, are defined as

\refstepcounter{equation}
$$
  \tilde u^f_i = \frac{\overline{(1-\phi)u^f_i}}{(1-\bar\phi)}, \quad 
  \tilde u^s_i = \frac{\overline{\phi u^s_i}}{\bar\phi},
  \eqno{(\theequation{\mathit{a},\mathit{b}})}
$$
with $\phi$ the solid phase volumetric concentration and $u^{f''}_i = u^f_i - \tilde u^f_i$ and $u^{s''}_i = u^s_i - \tilde u^s_i$ are the sub-grid scale velocity fluctuations.

The filtered two-phase flow equations are composed of the filtered fluid and solid phase continuity equations (\ref{cont_flu}) and  (\ref{cont_sol}) and the filtered fluid and solid phase momentum equations (\ref{mom_flu}) and (\ref{mom_sol}),

\begin{equation}
 \frac{\partial (1-\bar\phi)}{\partial t} + \frac{\partial (1- \bar\phi)\tilde u_i^f}{\partial x_i} = 0,
 \label{cont_flu}
\end{equation}


\begin{equation}
 \frac{\partial \bar\phi}{\partial t} + \frac{\partial\bar\phi\tilde u_i^s}{\partial x_i} = 0,
 \label{cont_sol}
\end{equation}


\begin{eqnarray}
\frac{\partial\rho^f(1-\bar\phi)\tilde u^f_i}{\partial t} + \frac{\partial\rho^f(1-\bar\phi)\tilde u^f_i \tilde u^f_j}{\partial x_j} = -(1-\bar\phi)\frac{\partial \bar P^f}{\partial x_i} + \frac{\partial}{\partial x_j}\left(\tilde T^f_{ij} + \sigma^{f, sgs}_{ij}\right) + \bar I_i
\nonumber\\ + \rho^f(1-\bar\phi)g_i + \Phi^{f,sgs}_i,
\label{mom_flu}
\end{eqnarray}


\begin{eqnarray}
\frac{\partial\rho^s\bar\phi\tilde u^s_i}{\partial t} + \frac{\partial\rho^s\bar\phi\tilde u^s_i \tilde u^s_j}{\partial x_j} = -\bar\phi\frac{\partial \bar P^f}{\partial x_i} - \frac{\partial \bar P^s}{\partial x_i} + \frac{\partial}{\partial x_j}\left(\tilde T^s_{ij} + \sigma^{s, sgs}_{ij}\right) - \bar I_i + \rho^s\bar\phi g_i 
\nonumber\\ + \Phi^{s,sgs}_i,
\label{mom_sol}
\end{eqnarray}
with $\rho^f$ and $\rho^s$ the fluid and solid densities, $g_i$ the acceleration of gravity, $\bar P^f$ and $\bar P^s$ the filtered fluid and solid pressures, $\tilde T^f_{ij}$ and $\tilde T^s_{ij}$ the filtered fluid and solid phase shear stress tensors, $\sigma^{f,sgs}_{ij}$, $\sigma^{s,sgs}_{ij}$, $\Phi^{f,sgs}_i$ and $\Phi^{s,sgs}_i$ the fluid and solid sub-grid scale stress tensors and other sub-grid scale contributions respectively presented in \S\ref{sgsmodel} and $\bar I_i$ the filtered momentum exchange term between the two phases.

The filtered fluid phase shear stress tensor is defined as:
\begin{equation}
 \tilde T^f_{ij} = \rho^f(1-\bar\phi)\nu^f\left(\frac{\partial\tilde u^f_i}{\partial x_j} + \frac{\partial\tilde u^f_j}{\partial x_i} - \frac{2}{3}\frac{\partial\tilde u^f_k}{\partial x_k}\delta_{ij}\right),
\end{equation}
with $\nu^f$ the fluid viscosity and $\delta_{ij}$ the Kronecker symbol. The filtered solid phase pressure $\bar P^s$ and shear stress tensor $\tilde T^s_{ij}$ are calculated using the kinetic theory for granular flows as detailed in \S\ref{kinetic_theory}.

The filtered momentum exchange term $\bar I_i$ between the two phases is composed of the drag, lift and added mass forces $\bar D_i$, $\bar L_i$ and $\bar A_i$, respectively following the expression

\begin{equation}
 \bar I_i = \bar D_i + \bar L_i + \bar A_i \quad \mbox{with\ }\quad \left\{
 \begin{array}{lll}
      \bar D_i = \frac{\rho^s\bar\phi}{\tilde{t_s}}\left(\tilde u^f_i - \tilde u^s_i\right) \\[2pt]
      \bar L_i = \bar\phi(1-\bar\phi)C_l\rho^m\Vert\tilde u^f_i - \tilde u^s_i\Vert\epsilon_{ijk}\frac{\partial\tilde u^m_k}{\partial x_j} \\[2pt]
      \bar A_i = \bar\phi(1-\bar\phi)C_a\rho^f\left[\frac{\partial\tilde u^f_i}{\partial t} + \frac{\partial\tilde u^f_i \tilde u^f_j}{\partial x_j} - \frac{\partial\tilde u^s_i}{\partial t} + \frac{\partial\tilde u^s_i \tilde u^s_j}{\partial x_j}\right]
  \end{array} \right.
 \label{mom_ex}
\end{equation}
where $C_l=0.5$ and $C_a=0.5$ are the lift and added mass coefficients, $\rho^m=\bar\phi\rho^s + (1-\bar\phi)\rho^f$ is the volume-averaged mixture density,  $\tilde u^m_i = \bar\phi\tilde u^s_i + (1-\bar\phi)\tilde u^f_i$ is the mixture velocity and $\tilde t_s$ is the particle response time following the drag law proposed by \cite{gidaspow1986}, \textit{i.e.}
\begin{equation}
 \tilde t_s = \frac{4}{3}\frac{\rho^s}{\rho^f}\frac{d_p}{C_D\Vert\tilde u^f_i - \tilde u^s_i\Vert}(1-\bar\phi)^{2.65} \quad \mbox{with\ }\quad \left\{
 \begin{array}{ll}
      C_D = \frac{24}{\Rey_p}\left(1+0.15\Rey_p^{0.687}\right) \\[2pt]
      \Rey_p = \frac{d_p\Vert\tilde u^f_i - \tilde u^s_i\Vert}{\nu^f}
  \end{array} \right.
 \label{drag}
\end{equation}
with $C_D$ the drag coefficient from \cite{schiller1933}.

\subsection{Sub-grid scale modelling}
\label{sgsmodel}
As a direct result of the filtering of the two-phase flow equations, additional sub-grid terms appear in the momentum equations. The fluid and solid phase sub-grid stress tensors $\sigma^{f,sgs}_{ij}=\rho^f(1-\bar\phi)(\widetilde{u^f_iu^f_j} - \tilde u^f_i \tilde u^f_j)$ and $\sigma^{s,sgs}_{ij}=\rho^s\bar\phi(\widetilde{u^s_iu^s_j} - \tilde u^s_i \tilde u^s_j)$ come from the filtering of the nonlinear advection terms in the momentum equations. Whereas \cite{cheng2018} modelled the sub-grid stress tensors using the dynamic procedure proposed by \cite{germano1991} and \cite{lilly1992}, in the present contribution they are modelled using the dynamic Lagrangian procedure proposed by \cite{meneveau1996}. The main difference is that model coefficients are averaged over streamlines (details can be found in appendix \ref{dynlag}) and not plane-averaged over homogeneous flow directions. The dynamic Lagrangian procedure has the advantage of getting rid of the necessity to have homogeneous directions and of preserving a certain locality in space making it applicable to more complex geometries and inhomogeneous flows in future research \citep{meneveau1996}. The sub-grid stress tensors are written as
\begin{equation}
 \sigma^{f,sgs}_{ij} = 2\rho^f(1-\bar\phi)\Delta^2 \vert \tilde{\boldsymbol S}^f \vert \left(C_1^f\tilde S_{ij}^f - \frac{1}{3}C_2^f\tilde S_{kk}^f\right),
\end{equation}
and
\begin{equation}
 \sigma^{s,sgs}_{ij} = 2\rho^s\overline\phi\Delta^2 \vert \tilde{\boldsymbol S}^s \vert\left(C_1^s\tilde S_{ij}^s - \frac{1}{3}C_2^s \tilde S_{kk}^s\right),
\end{equation}
with $\tilde S^f_{ij}$ and $\tilde S^s_{ij}$ the fluid and solid resolved strain rate tensor, respectively, and $C_1^f$, $C_2^f$, $C_1^s$, $C_2^s$ the dynamically computed model coefficients (details in appendix \ref{dynlag}). 

Other Eulerian-Eulerian sub-grid contributions resulting from the filtering of the pressure, stress and momentum exchange terms are represented by $\Phi^{f,sgs}_i$ and $\Phi^{s,sgs}_i$. These sub-grid terms are taking into account the effect of unresolved particle clusters and streamers having length scales smaller than the filter width $\Delta$. \cite{cheng2018} modelled the sub-grid momentum exchange term using a drift velocity model proposed by \cite{ozel2013}, but since the typical size of the smallest mesoscale structures is of the order of 10 to 100 particle diameters \citep{agrawal2001}, the sub-grid terms taking into account these effects should vanish for filter sizes of the order of the particle size. This has been confirmed by \cite{ozel2013} whom quantitatively reported the relative importance of sub-grid terms by explicitly filtering two-phase Eulerian-Eulerian DNS results for different filter sizes. In all the simulations presented in this paper, $\Delta$ is always of the order of the particle size or smaller and therefore, the sub-grid contributions $\Phi^{f,sgs}_i$ and $\Phi^{s,sgs}_i$ can be considered as negligible.

\subsection{Particle stress modelling}
\label{kinetic_theory}
For solid particles in a boundary layer flow, the solid phase volume fraction changes by several orders of magnitudes from the outer part of the flow to the bottom boundary. Therefore, the modelling methodology used to describe the disperse phase hydrodynamic has to be valid for a wide range of volume fractions from the dilute limit where the interaction with the carrier phase is dominant to higher volume fractions and collision-dominated regimes.

The filtered solid phase pressure $\bar P^s$ and shear stress tensor $\tilde T^s_{ij}$ are given by 
\begin{equation}
 \bar P^s = \rho^s\bar\phi\left[1 + 2(1+e_c)\bar\phi g_{s0}\right]\bar\Theta - \rho^s\lambda\frac{\partial\tilde u^s_k}{\partial x_k}\delta_{ij}
 \label{partpres}
\end{equation}
and 
\begin{equation}
\label{partstress}
 \tilde T^s_{ij} = \rho^s\bar\phi\nu^s\left(\frac{\partial\tilde u^s_i}{\partial x_j} + \frac{\partial\tilde u^s_j}{\partial x_i} - \frac{2}{3}\frac{\partial\tilde u^s_k}{\partial x_k}\delta_{ij}\right),
\end{equation}
respectively, from \cite{gidaspow1994} with the particle phase shear viscosity $\nu^s$, bulk viscosity $\lambda$ given by
\begin{equation}
\label{shearvisc}
 \nu^s = d_p\sqrt{\bar\Theta}\left[\frac{4\bar\phi^2g_{s0}(1+e_c)}{\sqrt{5\pi}} + \frac{\sqrt{\pi}g_{s0}(1+e_c)^2(2e_c-1)\bar\phi^2}{15(3-e_c)} + \frac{\sqrt{\pi}\bar\phi}{6(3-e_c)}\right]
\end{equation}
and 
\begin{equation}
\label{bulkvisc}
 \lambda = \frac{4}{3}\bar\phi^2\rho^sd_pg_{s0}(1+e_c)\sqrt{\frac{\bar\Theta}{\pi}}
\end{equation}
respectively, where $g_{s0} = (2-\bar\phi)/2(1-\bar\phi)^3$ the radial distribution function for dense rigid spherical particle gases from \cite{carnahan1969}, $e_c=0.8$ is the restitution coefficient for binary collisions and $\bar\Theta$ is the filtered granular temperature. According to \cite{fevrier2005}, $\bar\Theta$ represents the pseudo-thermal kinetic energy associated with the uncorrelated random motions of the particles and should not be confused with the turbulent sub-grid scale turbulent kinetic energy $k^s$ associated with correlated motions of the particles.

The filtered granular temperature $\bar\Theta$ is obtained by solving the following transport equation:
\begin{equation}
\label{kt}
 \frac{3}{2}\left[\frac{\partial\bar\phi\rho^s\bar\Theta}{\partial t} + \frac{\partial\bar\phi\rho^s\tilde u^s_i\bar\Theta}{\partial x_j}\right] = \Pi_R + \Pi_q + J_{int} - \gamma + \Phi^{sgs}_\Theta. 
\end{equation}
Here $\Pi_R$ the production of granular temperature by resolved flow scales given by
\begin{equation}
\label{granprod}
 \Pi_R = \left(-\bar P^s\delta{ij} + \tilde T^s_{ij}\right)\frac{\partial \tilde u^s_i}{\partial x_j},
\end{equation}
$\Pi_q$ the divergence of the granular temperature flux analogous to the Fourier's law of conduction given by
\begin{equation}
\label{granflux}
 \Pi_q = \dfrac{\partial}{\partial x_i}\left[-D_\Theta\frac{\partial\bar\Theta}{\partial x_j}\right],
\end{equation}
where $D_\Theta$ is the conductivity of the granular temperature calculated following 
\begin{equation}
\label{diffcoeff}
 D_\Theta = \rho^sd_p\sqrt{\bar\Theta}\left[\frac{2\bar\phi^2g_{s0}(1+e_c)}{\sqrt{\pi}} + \frac{9\sqrt{\pi}g_{s0}(1+e_c)^2(2e_c-1)\bar\phi^2}{2(49-33e_c)} + \frac{5\sqrt{\pi}\bar\phi}{2(49-33e_c)}\right],
\end{equation}
$\gamma$ is the dissipation rate of granular temperature given by
\begin{equation}
 \gamma = 3(1-e_c^2)\phi^2\rho^sg_{s0}\bar\Theta\left[\frac{4}{d_p}\sqrt{\frac{\Theta}{\pi}}-\frac{\partial u^s_j}{\partial x_j}\right],
 \label{thetadiss}
\end{equation}
$J_{int}$ is the fluid-particle interaction term and $\Phi^{sgs}_\Theta$ is the sub-grid term.

The fluid-particle interaction term presented in equation (\ref{inter}) represents the balance between the production of granular temperature due to the fluid pseudo-thermal kinetic energy $\bar\Theta^f$ and the dissipation due to drag. The formulation is similar to the interaction term presented in \cite{fox2014} but transcripted in the LES formalism with the difference that, for finite-size particles, the fluid pseudo-thermal kinetic energy is confounded with the sub-grid TKE. It is given by $\Theta^f = \frac{2}{3}\alpha k^f$, with the sub-grid fluid TKE $k^f=(C_1^f/C_\varepsilon)^{2/3}\vert \tilde{\boldsymbol S}^f \vert$, where $C_\varepsilon$ is an empirical constant of the order of unity coming from the Kolmogorov theory \citep{yoshizawa1985} taken to be $C_\varepsilon=1.048$ in recent studies \citep{arshad2019, chatzimichailidis2019, ries2020} and $\alpha=e^{-BSt}$ is a coefficient characterizing the degree of correlation between particles and fluid velocity fluctuations \citep{hsu2004}. The empirical parameter B is a tuning coefficient for Reynolds average models set to 1 in the present simulations.

\begin{equation}
 J_{int} = 3\frac{\rho^s\bar\phi}{\tilde t_s}(\bar\Theta^f - \bar\Theta)
 \label{inter}
\end{equation}

\begin{figure}
  \centerline{\includegraphics[width=0.8\textwidth]{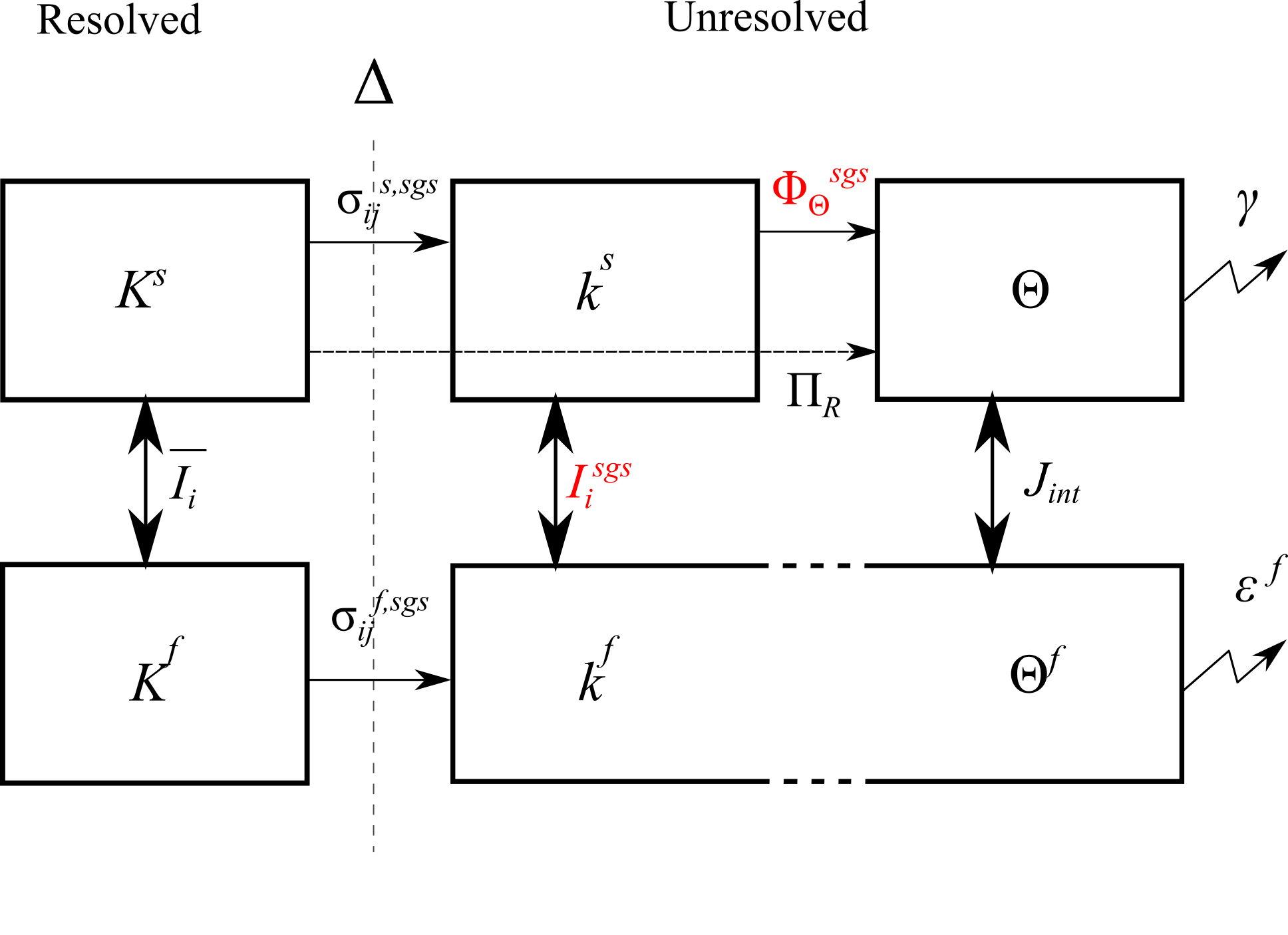}}
  \caption{Schematic representation of the energy transfers between fluid and solid resolved, unresolved, correlated and uncorrelated scales of the flow with $K^f$ and $K^s$ the resolved fluid and solid TKE, $k^f$ and $k^s$ the fluid and solid sub-grid TKE, $\Theta^f$ the fluid pseudo-thermal kinetic energy and $\Theta$ the granular temperature. Terms in red are neglected because the grid size $\Delta$ is of the order of the particles diameter $d_p$ \citep{agrawal2001, ozel2013}}
\label{transfkin}
\end{figure}

The sub-grid term $\Phi^{sgs}_\Theta=\rho^s\phi\varepsilon^s$, where $\varepsilon^s$ is the dissipation rate of solid sub-grid TKE represents the production of granular temperature from the energy transfer between the correlated solid sub-grid TKE and the granular temperature. Similarly to the sub-grid terms in the momentum equation, $\Phi^{sgs}_\Theta$ should vanish for grid sizes of the order of the particle diameter \citep{agrawal2001, ozel2013} and is therefore neglected. A summary of the energy transfers between fluid and solid resolved, unresolved, correlated and uncorrelated scales of the flow is presented in figure \ref{transfkin}.

\subsection{Finite-size correction model}
\label{fs_model}

\begin{figure}
  \centerline{\includegraphics[width=0.6\textwidth]{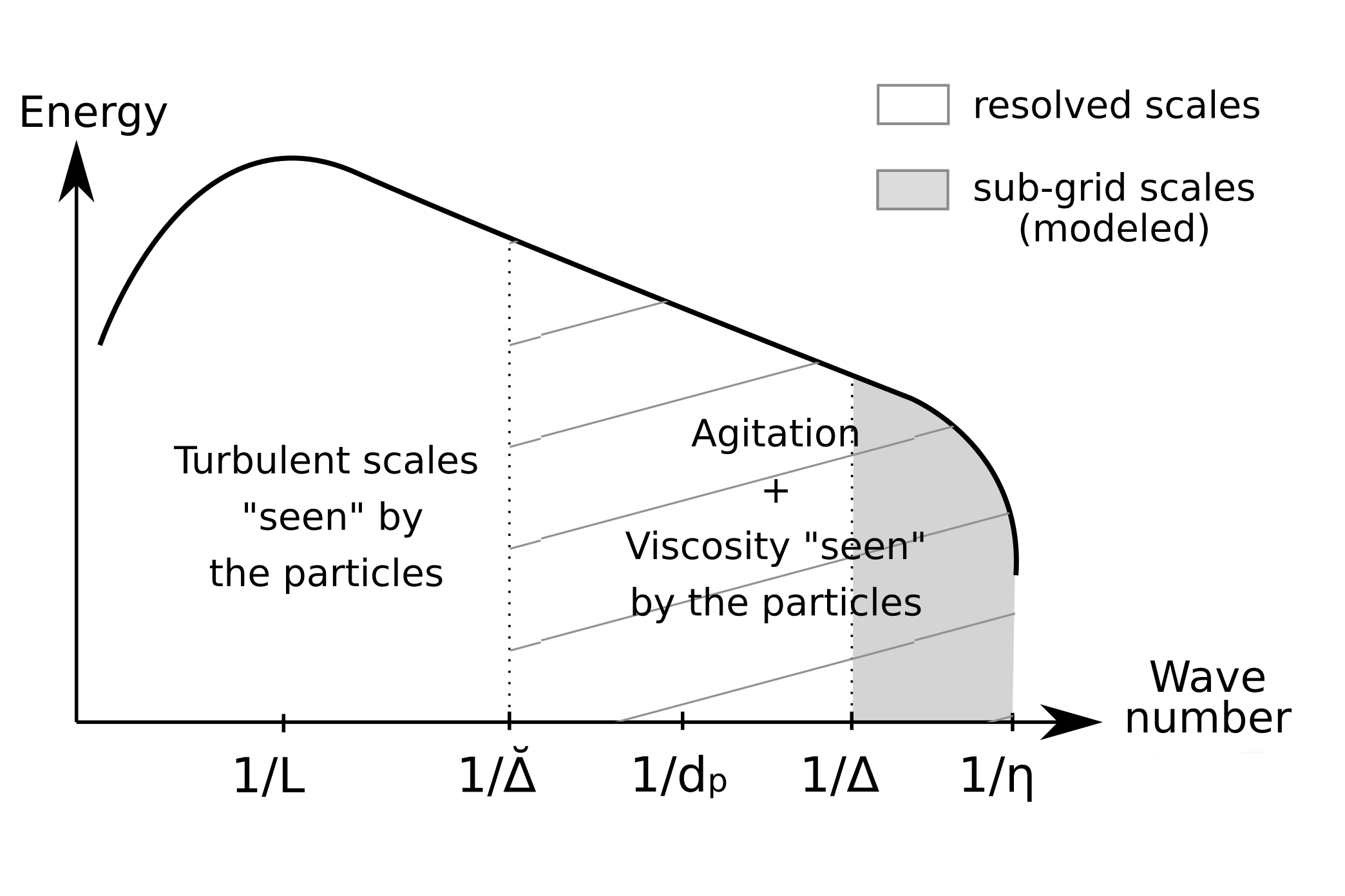}}
  \caption{Schematic representation of an idealized turbulent spectrum including the different flow scales and their contributions to the particles dynamics ($\eta$: Kolmogorov scale, $\Delta$: filter width, $d_p$: particle diameter, $\breve\Delta$: second filter width, $L$: integral scale of turbulence).}
\label{spectrum}
\end{figure}

Compared with particles smaller than the Kolmogorov length scale, finite-size particles do not only act as a temporal filter of the turbulent flow scales through the drag force but also as a spatial filter \citep{qureshi2007, calzavarini2009}. In order to take into account the finite-size effect of the particles in the Eulerian-Eulerian two-phase flow model, a distinction is made between turbulent eddies having larger or smaller length scales than the particle diameter $d_p$ (blank and hatched zones, respectively, of the idealized turbulent spectrum represented in figure \ref{spectrum}). Following observations made by \cite{qureshi2007} and \cite{calzavarini2009}, turbulent eddies larger than the particle diameter will contribute to the relative velocity between the two phases in the drag force as fluid velocity ``seen'' by the particles whereas smaller eddies are assumed to (1) modify the particle response time by increasing the viscosity ``seen'' by the particles by defining an effective turbulent viscosity at the particle scale following \cite{gorokhovski2018} and (2) contribute to particle agitation by increasing the production of granular temperature to be consistent with the energy transfers between correlated and uncorrelated solid phase velocity fluctuations \citep{fevrier2005,fox2014}. The idealized turbulent spectrum presented in figure \ref{spectrum} summarizes the different contributions of the turbulent fluid flow scales to the solid phase dynamics.

To take into account finite-size effects, the filtered effective drag force is rewritten as
\begin{equation} 
 \bar D_i = \frac{\rho^s \bar\phi}{\breve t_s}(\breve u^f_i - \tilde u^s_i),
\end{equation}
with $\breve u^f_i$ the fluid velocity ``seen'' by the particles corresponding to the resolved fluid phase velocity $\tilde u^f_i$ filtered at a scale $\breve\Delta \sim O(d_p)$. According to \cite{kidanemariam2013} the value of $\breve\Delta$ should not be too large to still be relevant to predict the particles motion but not too small to be sufficiently free from the local flow disturbances generated by the presence of the particles. To be able to determine the filter length $\breve\Delta$, they reported the ratio between the averaged magnitude of the flow velocity around spheres and the undisturbed flow field as a function of the distance from the centre of the sphere for different particle Reynolds numbers. From their analysis, around $80\%$ of the undisturbed mean flow velocity is recovered with a filter width taken as twice the diameter of the particle. Therefore, to compute the fluid velocity ``seen'' by the particles, the filter size is first chosen to be $\breve\Delta =  2 d_p$. A sensitivity analysis to the filter size is presented in \S\ref{sens_test}.

Whereas the turbulent scales smaller than the particle diameter are usually unresolved, due to the mesh refinement close to the wall, these turbulent scales are composed of both resolved and unresolved eddies in this region. In the present configuration, $\breve\Delta=\Delta_{x,z}$ in the streamwise and spanwise directions but $\breve\Delta>\Delta_y$ in the wall-normal direction. To calculate $\breve u^f_i$, a weighted average of the resolved fluid velocity in the wall-normal direction is performed using a Gaussian distribution with standard deviation $\breve\Delta$ to compute the weighting coefficients.

The new particle response time $\breve t_s$ still follows the drag law given by equation (\ref{drag}) but the relative velocity between the two phases is calculated using the filtered fluid velocity $\breve u^f_i$, and the expression for the particle Reynolds number is modified according to \cite{gorokhovski2018} to take into account the effect of turbulent scales smaller than the particles by the mean of a turbulent eddy viscosity $\nu^t_p$ at the scale of the particles following
\begin{equation}
 \Rey_p = \frac{d_p\Vert\breve u^f_i - \tilde u^s_i\Vert}{\nu^f+\nu^t_p}.
\end{equation}

The turbulent viscosity at the particle scale can be calculated using Kolmogorov scaling and Prandtl's mixing length hypothesis following $\nu^t_p \sim \varepsilon_p^{1/3}d_p^{4/3}$, with $\varepsilon_p$ the dissipation of TKE at the particle scale \citep{gorokhovski2018}. By assuming that the turbulent scales between $\breve\Delta$ and $d_p$ are in the inertial range of the turbulent spectrum, the approximation $\varepsilon_{\breve\Delta} = \varepsilon_{p}$ can be made with $\varepsilon_{\breve\Delta}$ the dissipation rate at the filter scale.

The expression of the dissipation rate at the filter scale $\varepsilon_{\breve\Delta}$ is estimated following the expression from \cite{yoshizawa1985} defined as a function of the filter width $\breve\Delta$ and the total TKE below $\breve\Delta$ defined as the sum of $\breve k=\frac{1}{2}\tilde u^{f''}_i\tilde u^{f''}_i$ the resolved TKE (from $\breve\Delta$ to $\Delta$), with $\tilde u^{f''}_i=\tilde u^f_i - \breve u^f_i$ and $k^f$ the sub-grid TKE (from $\Delta$ to $\eta$), \textit{i.e.}
\begin{equation}
 \varepsilon_{\breve\Delta} = C_\varepsilon\frac{(\breve k + k^f)^{3/2}}{\breve\Delta}.
\label{diss_FS}
 \end{equation}

Eventually, the particle response time with finite-size correction is written as
\begin{equation}
 \breve t_s = \frac{4}{3}\frac{\rho^s}{\rho^f}\frac{d_p}{C_D\Vert\breve u^f_i - \tilde u^s_i\Vert}(1-\bar\phi)^{2.65} \quad \mbox{with\ }\quad \left\{
 \begin{array}{ll}
      C_D = \frac{24}{\Rey_p}\left(1+0.15\Rey_p^{0.687}\right) \\[2pt]
      \Rey_p = \frac{d_p\Vert\breve u^f_i - \tilde u^s_i\Vert}{\nu^f+\varepsilon_{\breve\Delta}^{1/3}d_p^{4/3}}
  \end{array} \right .
 \label{dragfs}
\end{equation}

Furthermore, the turbulent flow scales below $\breve\Delta$ contribute to increase the production of granular temperature isotropically. The fluid particle-interaction term $J_{int}$ in equation (\ref{kt}) includes the resolved sub-particle TKE, \textit{i.e.}

\begin{equation}
\label{jintfs}
 J_{int} = \frac{\rho^s}{\breve t_s}\frac{\phi}{1-\phi}\left[2\alpha (\breve k+k^f) - 3\bar\Theta\right].
\end{equation}

It shall be mentioned that the proposed model tends to the two-fluid model in its traditional formulation for particles smaller than the Kolmogorov length scale ($d_p/\eta<1$). Indeed, if $\breve\Delta\leq\eta$ then $\breve u^f_i=\tilde u^f_i$ and, therefore, $\breve k = k^f =0$. The turbulent viscosity at the particle scale vanishes $\nu^t_p=0$ and eventually $\breve t_s = \tilde t_s$. Furthermore, the proposed finite-size correction model allows us to approach the exact solution for vanishingly small mesh sizes. Indeed, the correction model allows us to decorrelate the filter width associated with the particle size in the drag law and the filter size imposed by the mesh for the LES making the solution mesh independent.

\subsection{Numerical implementation}

The present model is adapted from the turbulence averaged two-phase flow solver sedFoam (https://github.com/sedFoam/sedFoam) \citep{cheng2017, chauchat2017}. It is implemented in the open-source computational fluid dynamics toolbox OpenFoam \citep{jasak2020} and solves the Eulerian-Eulerian two-phase flow mass and momentum equations using a finite volume method and a pressure-implicit with splitting of operators (PISO) algorithm for velocity-pressure coupling \citep{rusche2002}. In the PISO algorithm, at each time step, intermediate velocities are first computed by solving the momentum equations without the pressure gradient term. Then, the Poisson equation for the pressure is solved in order to calculate the corrected pressure field and ensure mass conservation. Eventually, the velocity fields are corrected based on the new pressure field. Several steps can be applied to the velocity prediction-correction to increase convergence ({\it nCorrectors} in OpenFoam). In the present simulations, {\it nCorrectors} = 2 is sufficient for convergence. More information about consistency, algorithm and numerical implementation can be found in \cite{chauchat2017}.

In the present simulations the same numerical schemes as in \cite{cheng2018} are used to provide a second-order accuracy in both space and time. A second-order implicit backward scheme is used for temporal derivatives (denoted as {\it backward} in OpenFoam) and a second-order total variation diminish (TVD) scheme is used for the mass conservation equation and the granular temperature transport equation (denoted as {\it limitedLinear} in OpenFoam). For the advection terms in the momentum equations, a second-order centered scheme is used for which high frequency filtering of the oscillations induced by second-order discretization is performed by introducing a small amount of upwind scheme (denoted as {\it filteredLinear} in OpenFoam). The gradient are computed using a second-order centered scheme (denoted as {\it linear} in OpenFoam).

%
%
\section{Results}

In this section numerical simulations performed on different flow configurations are presented in order to assess the two-fluid LES model presented in \S2. First, a clear water configuration, {\it i.e.} without particles ($\bar\phi=0$), is presented and compared with existing experimental and numerical DNS data to validate the model, the choice of the grid resolution and the numerical schemes. Second, three particle-laden flow configurations involving finite-sized particles are reproduced numerically to evaluate the capability of the two-fluid LES model, including the finite-size correction model, to predict the turbulent suspension of particles.

\subsection{Clear water configuration}
\label{CW}
The clear water configuration from \cite{kiger2002} consists of a closed unidirectional channel flow with Reynolds number $\Rey_\tau=u_\tau h/\nu^f=560$ based on the wall-friction velocity $u_\tau = 2.8\times 10^{-2} \ m.s^{-1}$ and channel half-height $h=0.02 \ m$.

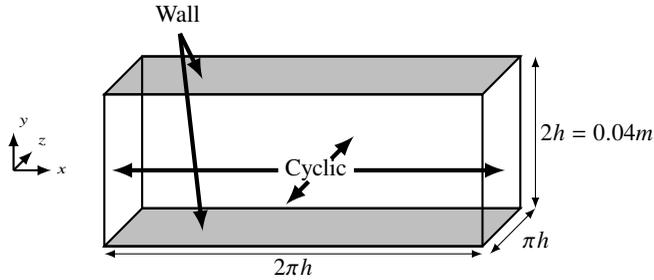
\begin{figure}
\begin{center}
\begin{tikzpicture}[scale=1]
\tiny
\tikzstyle{legend}   = [ultra thick,-latex]
\tikzstyle{normal}   = [thick]
\tikzstyle{stress} = [-latex]
\tikzstyle{dim}    = [latex-latex]After the two-phase flow model validation without particles, t
\tikzstyle{axis}   = [thick,-latex,black!100]
\fill [gray!50] [normal] (-3,-1)--(-2.5,-0.5)--(2.5,-0.5)--(2,-1)--cycle;
\fill [gray!50] [normal] (-3,1)--(-2.5,1.5)--(2.5,1.5)--(2,1)--cycle;
\draw [normal] (-3,-1)--(-3,1)--(2,1)--(2,-1)--cycle;
\draw [normal] (-2.5,-0.5)--(-2.5,1.5)--(2.5,1.5)--(2.5,-0.5)--cycle;
\draw [normal] (-3,-1)--(-2.5,-0.5)--(2.5,-0.5)--(2,-1)--cycle;
\draw [normal] (-3,1)--(-2.5,1.5)--(2.5,1.5)--(2,1)--cycle;
\draw[axis] (-4.2, 0) -- ++(0.5,0) node[right] {$x$};
\draw[axis] (-4.2, 0) -- ++(0,0.5) node[above right] {$y$};
\draw[axis] (-4.2, 0) -- ++(0.25,0.25) node[above right] {$z$};
\draw[dim] (2.7,-0.5) -- ++(0,2) node[midway,right] {\small$2h=0.04m$};
\draw[dim] (-3,-1.1) -- ++(5,0) node[midway,below] {\small$2\pi h$};
\draw[dim] (2.1,-1.1) -- ++(0.6,0.6) node[at end,below right, pos=0.6] {\small $\pi h$};
\draw (-0.2,0) node {\small Cyclic};
\draw[legend] (-0.7,0) -- ++(-2.2,0);
\draw[legend] (0.3,0) -- ++(2,0);
\draw[legend] (-0.3, -0.15) -- ++(-0.3,-0.3);
\draw[legend] ( 0, 0.15) -- ++(0.3,0.3);
\draw[legend] (-2, 1.8) -- ++(0.3,-0.6) node[above,pos=-0.1] {\small Wall};
\draw[legend] (-2, 1.8) -- ++(0.3,-2.6);
\end{tikzpicture}
\caption{\label{num_dom}Sketch of the geometry and boundary conditions of the numerical domain for the simulation of the clear water configuration and configuration GB from \citet{kiger2002}.}
\end{center}      
\end{figure}

The numerical domain is a bi-periodic rectangular box (figure \ref{num_dom}) with cyclic boundary conditions in $x$ and $z$ directions and no slip boundary condition at the top and bottom boundaries for the velocities. The gradient of any other quantities is set to zero at the walls. The flow is driven by a pressure gradient along the $x$-axis dynamically adjusted at each time step in order to match the experimental bulk velocity $U_b=0.51 \ m.s^{-1}$. The mesh is composed of $314\times220\times160$ elements corresponding to a total of $11,105,280$ cells. The spanwise and streamwise resolution is constant with $\Delta_x^+\approx\Delta_z^+ \approx 11$ wall units ($+$ symbol with $\psi^+ = \psi u_\tau/\nu^f$). The mesh is stretched along the $y$-axis with $\Delta_y^+\approx1$ at the wall and $\Delta_y^+ \approx 6$ at the centreline. The time step is fixed to $\Delta t=10^{-4} \ s$ to ensure a maximum Courant-Friedrichs-Lewy number (CFL) lower than $0.3$ for stability reasons. In a recent publication , \cite{montecchia2019} performed a sensitivity analysis to the CFL number (CFL=0.1, 0.2 and 0.3) and the results did not show strong sensitivities.

All the simulations presented in this paper are initialized with a fully developed turbulent boundary layer flow obtained from a preliminary simulation. A first run is conducted to let the turbulence develop until the wall-friction velocity and the integral of the total flow kinetic energy have reached a steady state. This corresponds to approximately $200 T_b$, with $T_b=h/U_b$ the non-dimensional bulk time scale of the flow. Then, a second run is performed to compute turbulence statistics and Favre-averaged quantities over a duration of $200 T_b$. The Favre-averaging procedure is represented by the operator $\langle \cdot \rangle_F$ (details can be found in appendix \ref{average}). In clear water flow conditions, Favre averaging is equivalent to ensemble averaging denoted as $\langle\cdot\rangle$.

Similarly to what has been done by \cite{kiger2002}, the average profiles obtained experimentally and numerically are compared with the profiles from the DNS of \cite{moser1999} with $\Rey_\tau=590$. Since the Reynolds number in the configuration from \cite{kiger2002} is close to the DNS, it is reasonable to compare the profiles between the two configurations.

\begin{figure}
  \centerline{\includegraphics[width=\textwidth]{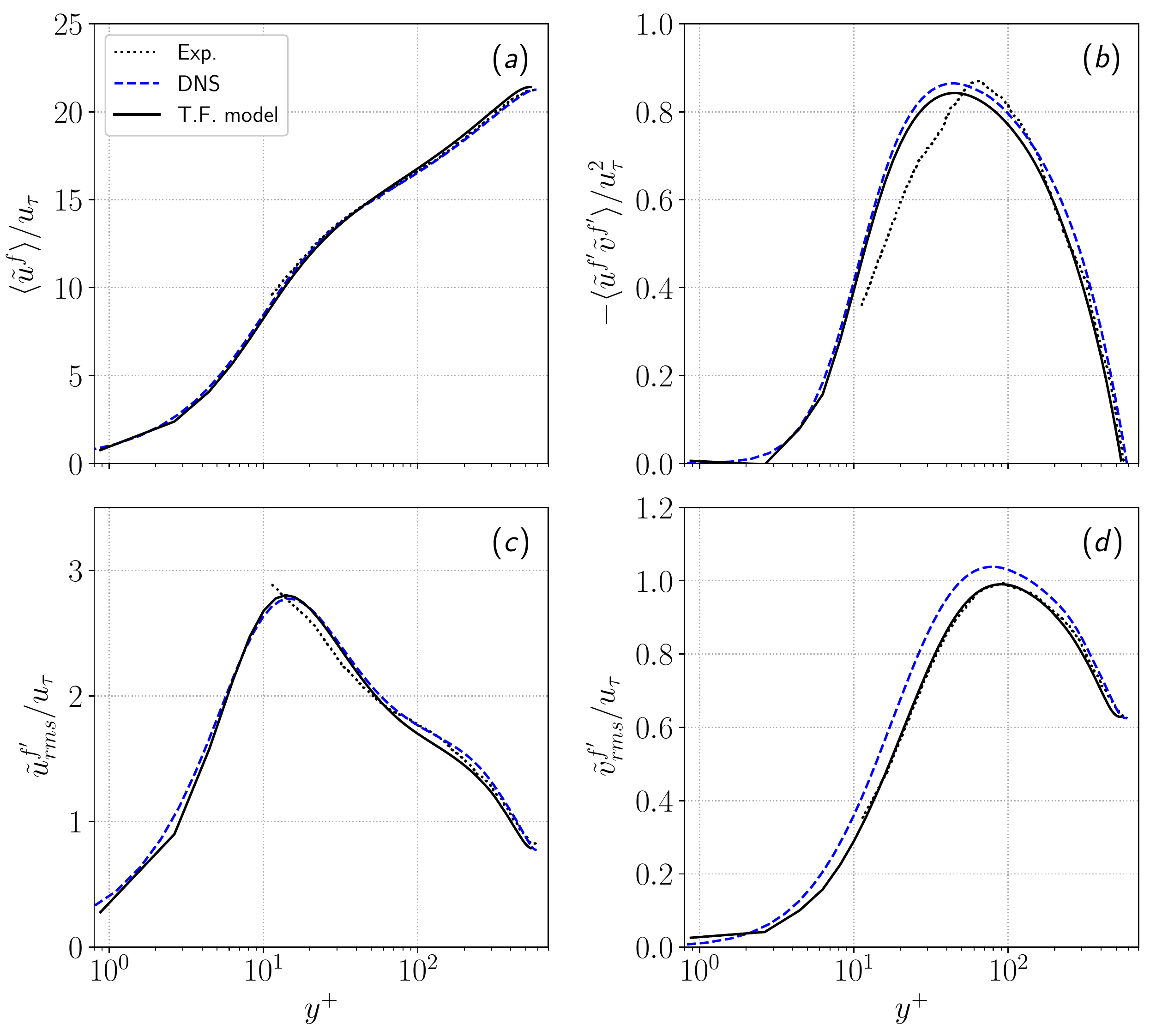}}
  \caption{Average profiles of velocity in $(a)$, Reynolds stress in $(b)$, r.m.s. of streamwise velocity fluctuations in $(c)$ and r.m.s of wall-normal velocity fluctuations in $(d)$ from the two-fluid model (T.F. model) compared with the numerical results from \cite{moser1999} (DNS) and experimental data from \cite{kiger2002} (Exp.).}
\label{vel_cw}
\end{figure}

The averaged velocity profiles, Reynolds stress and root-mean-square (r.m.s.) of the streamwise velocity fluctuations $\tilde u^{f'}_{rms}$ and wall-normal velocity fluctuations $\tilde v^{f'}_{rms}$ are presented in figure \ref{vel_cw} in wall units. In the simulations, the friction velocity is calculated based on the time-averaged streamwise pressure gradient as
\begin{equation}
\label{fricvel}
 u_\tau = \sqrt{h \Big\langle \frac{\partial \bar P^f}{\partial x}\Big\rangle}.
\end{equation}
The computed wall-friction Reynolds number is equal to $\Rey_\tau = 544$  ($u_\tau=2.72\times 10^{-2} \ m.s^{-1}$) which corresponds to an error below $3\%$ compared with the experiments.

The present clear water simulation produces profiles of averaged velocity and turbulence statistics that agree very well with the DNS and experimental data. However, especially for the Reynolds stress, some discrepancies between experimental measurements and the simulations appear near the wall. \cite{kiger2002} stated that their measurements can be considered highly reliable in the outer log layer with less than $5\%$ error for $y^+> 50$ and up to $25\%$ variability for $y^+ < 50$.

The agreement between numerical and experimental data confirms that without solid particles, the two-phase flow model behaves exactly as a single-phase flow model. The accurate prediction of the flow hydrodynamics and turbulent statistics allows us to validate the model implementation, the choices of numerical parameters and gives confidence to perform particle-laden simulations in the next sections.

\subsection{Particle-laden configurations}
\label{plconf}
In this section particle-laden configurations involving spherical glass beads (GB) from \cite{kiger2002}, natural sediment (NS) particles from \cite{muste2005} and almost neutrally buoyant sediment (NBS) particles from \cite{muste2005} are reproduced numerically. The flow and particles parameters are presented in table \ref{param} 

The targeted configurations correspond to the turbulent dilute suspended sediment transport boundary layer flows. In this situation, particles are entrained into suspension by the turbulent coherent flow structures and under steady-state flow conditions, an equilibrium concentration profile across the water depth establishes as the result of an equilibrium between the gravity driven settling flux $v_s\langle\phi\rangle$, with $v_s$ the settling velocity of the particles, and the turbulent Reynolds sediment flux $\langle v^{s'}\phi'\rangle$ \citep{rouse1938}, with $v^{s'}$ the solid phase vertical velocity fluctuations and $\phi'$ the sediment concentration fluctuations. By analogy with Fickian diffusion, this Reynolds sediment flux is modelled  using a gradient diffusion model. Introducing this model in the Reynolds-averaged sediment mass balance leads to the equation
\begin{equation}
  \label{rouse}
\displaystyle  v_s\langle\phi\rangle - \frac{\nu^f_t}{S_c}\frac{d\langle\phi\rangle}{dy} = 0,
\end{equation}
with $\nu^f_t$ the turbulent eddy viscosity (or turbulent momentum diffusivity) and $S_c$ the turbulent Schmidt number representing the efficiency of the sediment diffusion relative to $\nu_t^f$. For $S_c <1$, sediment particles are dispersed more efficiently by turbulence than fluid parcels. 

Given that $\nu^t_f=l_m^2 d\langle u^f \rangle/dy$ using Prandtl's mixing length $l_m=\kappa y$, with $\kappa=0.41$ the von Karman constant, and using the log-law-of-the-wall to describe the mean velocity profile, equation (\ref{rouse}) can be integrated analytically to give the following expression for the Reynolds-averaged particle concentration profile:
\begin{equation}
\displaystyle \frac{\langle\phi\rangle}{\phi_0} = \left(\frac{y_0}{y}\right)^{Ro}.
 \label{rouse2}
\end{equation}
Here $\phi_0$ is a reference concentration at a given reference elevation $y_0$ and $R_o=S_c v_s/u_\tau\kappa$ is the Rouse number. For open-channel flows, a free surface correction to the Prandtl's mixing length is introduced $l_m=\kappa y \sqrt{1-y/h}$, with $h$ representing the water depth, and the Reynolds-averaged particle concentration profile reads as
\begin{equation}  
\displaystyle \frac{\langle\phi\rangle}{\phi_0} =\left[\frac{y}{h-y}\frac{h-y_0}{y_0}  \right]^{-Ro}.
 \label{rouse2OC}
\end{equation}

 These two analytical solutions provide a reference with which the two-fluid LES model results can be compared. The value of $S_c$ is still debated in the sediment transport community \citep{lyn2008}, the most widely accepted model is the one from \cite{vanrijn1984} relating the turbulent Schmidt number to the suspension number, $S_c=(1+2(v_s/u_\tau)^2)^{-1}$. Nevertheless, a lot of scatter is observed on existing experimental data and no satisfactory explanation exists to support van Rijn's empirical formula \citep{lyn2008}.

The hydrodynamic configuration, numerical domain and parameters for configuration GB are the same as the clear water case presented in \S\ref{CW}. The only difference comes from the addition of a given amount of particles in the flow corresponding to a mean volumetric concentration of particles in the channel $\phi_{tot}=2.31\times 10^{-4}$. The particles are spherical and mono-dispersed with diameter $d_p=195 \ \mu m$ ($d_p^+\approx5.5$) and density $\rho^s = 2600 \ kg.m^{-3}$. For such particles, the computed fall velocity in still water using the drag law from equation (\ref{drag}) is $v_s=2.4\times10^{-2} \ m.s^{-1}$ ($v_s/u_\tau=0.85$). 

\begin{figure}
\begin{center}
\begin{tikzpicture}[scale=1]
\tiny
\tikzstyle{legend}   = [ultra thick,-latex]
\tikzstyle{normal}   = [thick]
\tikzstyle{stress} = [-latex]
\tikzstyle{dim}    = [latex-latex]
\tikzstyle{axis}   = [thick,-latex,black!100]
\fill [gray!50] [normal] (-3,-1)--(-2,0)--(3,0)--(2,-1)--cycle;
\draw [normal] (-3,-1)--(-3,0.5)--(2,0.5)--(2,-1)--cycle;
\draw [normal] (-2,0)--(-2,1.5)--(3,1.5)--(3,0)--cycle;
\draw [normal] (-3,-1)--(-2,0)--(3,0)--(2,-1)--cycle;
\draw [normal] (-3,0.5)--(-2,1.5)--(3,1.5)--(2,0.5)--cycle;
\draw[axis] (-4.2, 0) -- ++(0.5,0) node[right] {$x$};
\draw[axis] (-4.2, 0) -- ++(0,0.5) node[above right] {$y$};
\draw[axis] (-4.2, 0) -- ++(0.25,0.25) node[above right] {$z$};
\draw[dim] (3.2,0) -- ++(0,1.5) node[midway,right] {\small$h=0.021m$};
\draw[dim] (-3,-1.1) -- ++(5,0) node[midway,below] {\small$2\pi h$};
\draw[dim] (2.1,-1.1) -- ++(1.1,1.1) node[at end,below right, pos=0.6] {\small $\pi h$};
\draw (-0.2,0.2) node {\small Cyclic};
\draw[legend] (-0.8,0.2) -- ++(-2,0);
\draw[legend] (0.4,0.2) -- ++(2.1,0);
\draw[legend] (-0.2, 0.05) -- ++(-0.4,-0.4);
\draw[legend] ( 0.1, 0.35) -- ++(0.4,0.4);
\draw[legend] (-2, 1.8) -- ++(0.3,-0.6) node[above,pos=-0.1] {\small Symmetry plane};
\draw[legend] (-1.4, -0.4) -- ++(0.3,-0.4) node[above,pos=-0.1] {\small Wall};
\end{tikzpicture}
\caption{\label{num_dom_muste}Sketch of the geometry and boundary conditions of the numerical domain for configurations NS, NBS and NS* from \citet{muste2005}.}
\end{center}      
\end{figure}
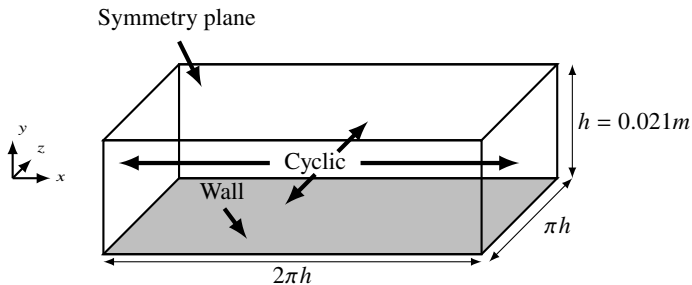

Configurations NS and NBS from \cite{muste2005} consist of turbulent particle-laden open-channel flows with water depth $h=0.021m$ in which finite-sized particles with density $\rho^s=2650 \ kg.m^{-3}$ and $\rho^s=1025 \ kg.m^{-3}$, respectively, are seeded. The NS and NBS hydrodynamic conditions are the same with a bulk velocity $U_b=0.84 \ m.s^{-1}$ and a targeted friction velocity $u_\tau=4.2\times10^{-2} \ m.s^{-1}$ corresponding to a Reynolds number based on the wall-friction velocity $\Rey_\tau=882$. Both types of particles have the same diameter $d_p=230 \ \mu m$ ($d_p^+\approx9.7$) resulting in a larger fall velocity ({\textit{ i.e.}} larger suspension number) for NS $v_s=2.4\times10^{-2} \ m.s^{-1}$ ($v_s/u_\tau=0.57$) compared with NBS $v_s=6\times10^{-4} \ m.s^{-1}$ ($v_s/u_\tau=0.01$). For both configurations, the mean volumetric concentration of sediment is equal to $\phi_{tot}=4.6\times 10^{-4}$. The computational domain is a rectangular box with bi-periodic boundary conditions along $x$- and $z$-axis and no slip boundary conditions at the bottom boundary (figure \ref{num_dom_muste}).

\begin{table}
  \begin{center}
\def~{\hphantom{0}}
  \begin{tabular}{lccccc}
    Parameters & ~~~~Units & ~~~~GB & ~~~~NS & ~~~~NBS & ~~~~NS* \\[3pt]
    $U_b$ & ~~~~$m.s^{-1}$ & ~~~~$0.51$ & ~~~~$0.84$ & ~~~~$0.84$ & ~~~~$0.84$ \\
    $u_\tau$($\times10^{-2}$) & ~~~~$m.s^{-1}$ & ~~~~$2.99$ &  ~~~~$4.20$ & ~~~~$4.20$ &  ~~~~$4.20$\\
    $h$ & ~$m$ & ~~~~$0.02$ & ~~~~$0.021$ & ~~~~$0.021$ & ~~~~$0.021$\\
    $\rho^s$ & ~~~~$kg.m^{-3}$ & ~~~~$2600$ & ~~~~$2650$ & ~~~~$1025$ & ~~~~$2650$ \\
    $d_p$ & ~~~~$\mu m$ & ~~~~$195$ & ~~~~$230$ & ~~~~$230$ & ~~~~$230$ \\
    $\phi_{tot}$($\times10^{-4}$) & ~~~~- & ~~~~$2.31$ & ~~~~$4.6$ & ~~~~$4.6$ & ~~~~$16.2$\\
    $v_s/u_\tau$ & ~~~~- & ~~~~$0.87$ &  ~~~~$0.54$ & ~~~~$0.01$ &  ~~~~$0.54$\\
    $\Rey_{p}$ & ~~~~- & ~~~~$4.8$ & ~~~~$9.1$ & ~~~~$0.39$ & ~~~~$9.1$\\
    $St$ & ~~~~- & ~~~~$3.2$ & ~~~~$5.7$ & ~~~~$6.6$ & ~~~~$5.7$\\
    $d_p/\eta$ & ~~~~- & ~~~~$5.5$ & ~~~~$9.7$ & ~~~~$9.7$ & ~~~~$9.7$\\

  \end{tabular}
  \caption{Flow and particles parameters for configurations GB, NS, NBS and NS*.}
  \label{param}
  \end{center}
\end{table}


The mesh is composed of $8,323,000$ cells with uniform streamwise and spanwise grid resolutions $\Delta_x^+ = \Delta_z^+ = 19$. The mesh resolution is stretched along the $y$-axis with the first grid point located at $\Delta_y^+ \approx 1$ and $\Delta_y^+ \approx 3$ at the top. Here $\Rey_p$, $St$ and $d_p/\eta$ are calculated based on the scaling analysis from \cite{finn2016}.

\begin{figure}
  \centerline{\includegraphics[width=0.7\textwidth]{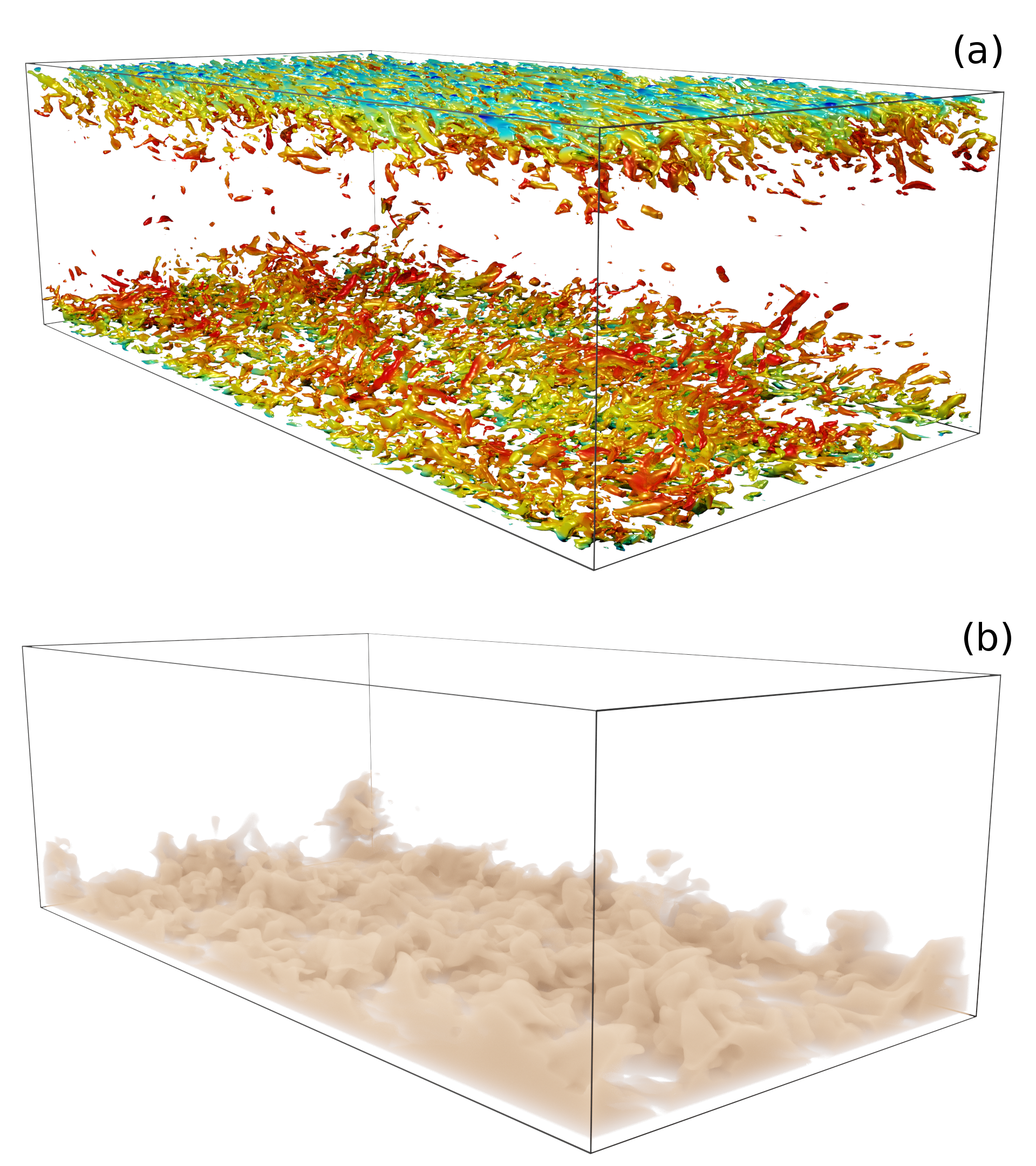}}
  \caption{Visualization of the instantaneous turbulent coherent structures using an isocontour of Q-criterion coloured by the velocity (panel (a)) and volume rendering of the concentration (panel (b)) from the GB configuration.}
\label{snapshot}
\end{figure}

\begin{figure}
  \centerline{\includegraphics[width=\textwidth]{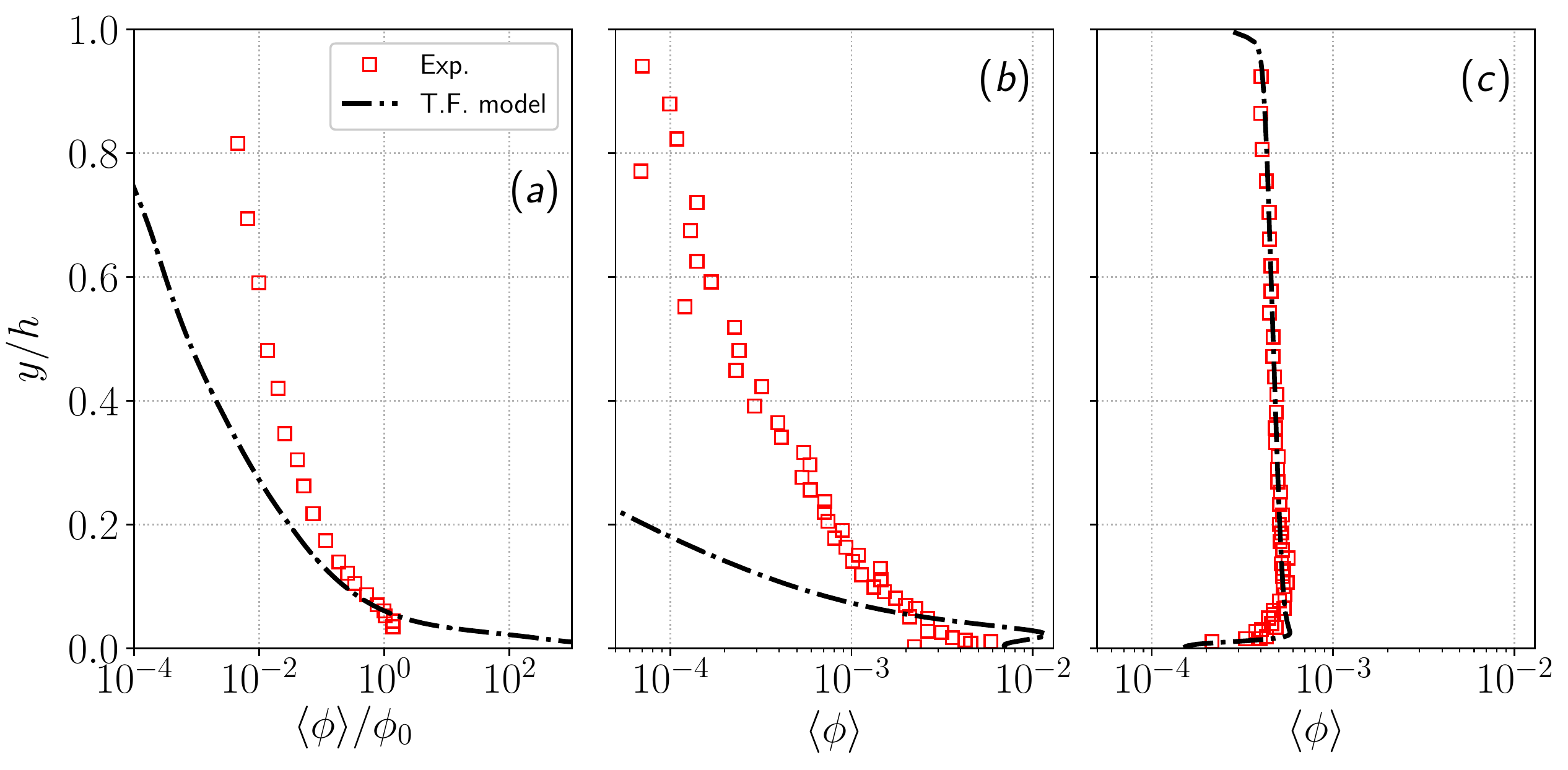}}
  \caption{Solid phase volumetric concentration profiles from the experiments (Exp.) and two-phase flow simulations (T.F. model) from configurations GB (panel (a)), NS (panel (b)) and NBS (panel (c)). In panel (a) experimental and numerical concentration profiles are normalized by the reference concentration $\phi_0$ taken at $y_0=0.06h$.}
\label{conc_NS_NBS}
\end{figure}

A first set of simulations for each configuration is performed in order to evaluate the predictive capability of the two-fluid model without finite-size correction. The visualization of the instantaneous turbulent coherent structures using an isocontour of Q-criterion (figure \ref{snapshot}a) and volume rendering of the concentration (figure \ref{snapshot}b) from the GB configuration shows the imprint of turbulence on the sediment concentration field. The differences of sediment concentration relative to the coherent structures highlights the importance of turbulence-particle interactions. The averaged solid phase concentration profiles obtained experimentally and numerically are compared in figure \ref{conc_NS_NBS}. For GB (figure \ref{conc_NS_NBS}a), experimental and numerical concentration profiles are normalized by the reference concentration $\phi_0$ taken at $y_0=0.06h$. For both configurations GB and NS (figure \ref{conc_NS_NBS}a and \ref{conc_NS_NBS}b), the volume fraction of particles in suspension is significantly under-estimated compared with the experimental data. However, for the NBS configuration (figure \ref{conc_NS_NBS}c), the average concentration profile predicted by the two-fluid model fits perfectly well the experimental results. For $v_s/u_\tau\ll1$, the weight of the particles is entirely supported by turbulence \citep{berzi2016}. The two-phase flow model in its original formulation correctly reproduces the vertical balance between settling and Reynolds fluxes. For this flow and these particle parameters, finite-size effects can be considered as negligible and the two-fluid model shows very good predictive capabilities without the finite-size correction model. In the following, configurations GB and NS for which the suspension number is higher are further investigated to understand the physical origin of the observed discrepancies. 

\begin{figure}
  \centerline{\includegraphics[width=\textwidth]{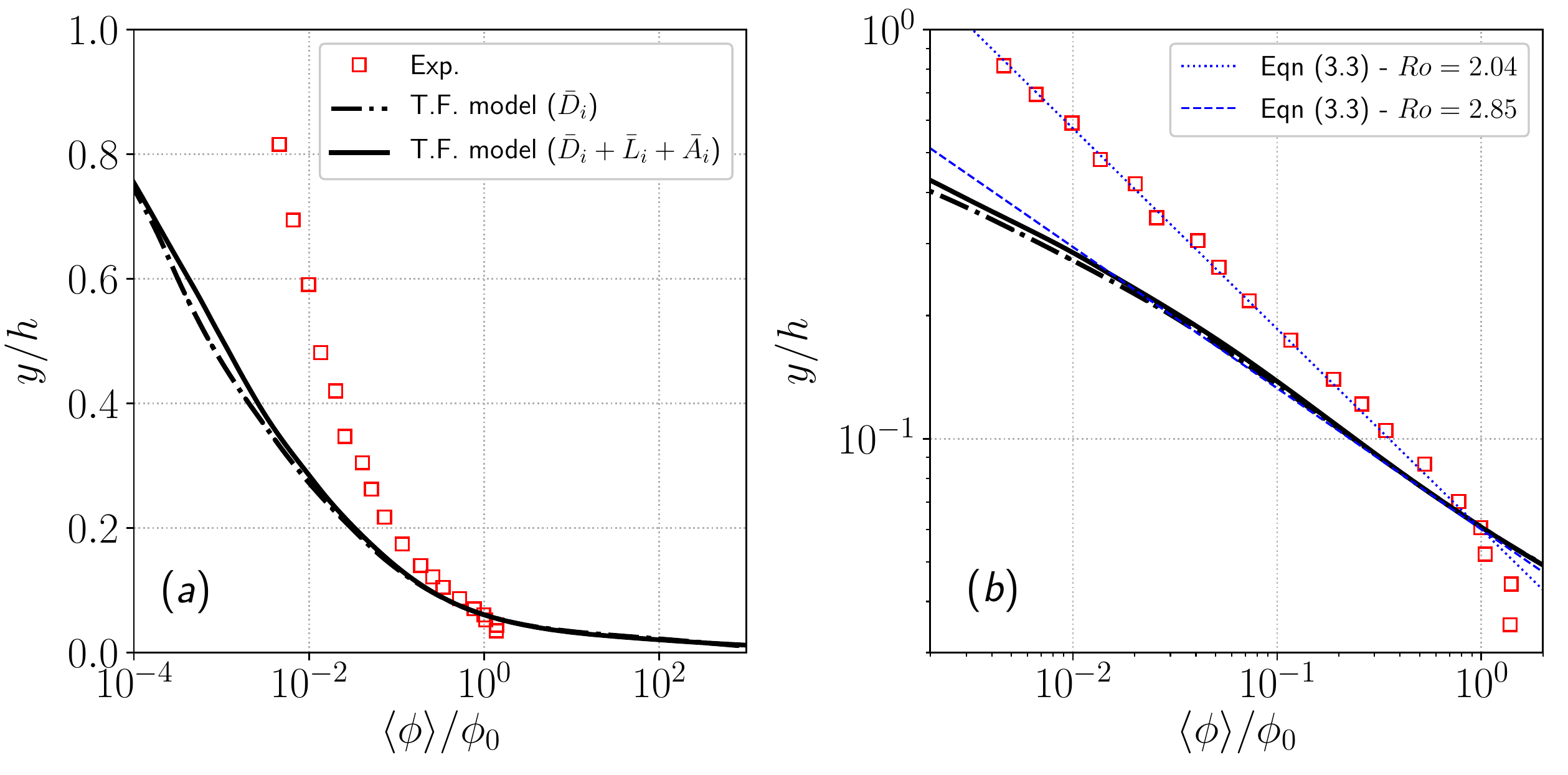}}
  \caption{Solid phase volumetric concentration profiles from the experiment (Exp.), two-phase flow simulation including only the drag force (T.F. model ($\bar D_i$)) and two-phase flow simulation including drag, lift and added mass forces (T.F. model ($\bar D_i + \bar L_i + \bar A_i$)) from configuration GB in semi-log scale in panel (a) and in log-log compared with analytical profiles from equation (\ref{rouse2}) with $Ro=2.04$ and $Ro=2.85$ in panel (b).}
\label{force_conc}
\end{figure}

The research hypothesis developed in this work is that the discrepancies observed in figure \ref{conc_NS_NBS} are due to finite-size effects. One could also argue that these discrepancies are due to missing fluid-particle forces such as added mass and lift forces. A simulation including these two forces have been performed for the configuration GB and the averaged concentration profiles are compared with the experiments and the analytical concentration profile from equation (\ref{rouse2}) in figure \ref{force_conc}. Since this expression is derived for an infinite boundary layer, one has to keep in mind that for closed channel flows, this expression could become less accurate near the centreline of the channel. The comparison between the simulation including only the drag force and the simulation including drag, lift and added mass forces indicates that the drag force is the dominant interaction force for this configuration. The contributions from lift and added-mass forces are almost negligible in this problem. The concentration profiles from both simulations show a power law that fits with (\ref{rouse2}) for $Ro=2.85$, whereas $Ro=2.04$ in the experiments (figure \ref{force_conc}b). The Basset history force does not appear in the momentum exchange term between the two phases since it is defined from a purely Lagrangian point of view. It would therefore be very difficult to obtain a volume average expression of the history force in the Eulerian formalism. To the best of the authors knowledge, there is no references in the literature showing the Eulerian expression of the Basset history force. The authors believe that the Basset history force would be significant very near the bottom boundary where wall-particle collisions occur but should not affect too much the vertical distribution of particles in the upper part of the channel where particle acceleration is weaker. It is only through a detailed comparison with Lagrangian point-particle simulations including the Basset history force that the role of this force could be investigated which is beyond the scope of the present work.

The two-phase flow model in its initial formulation, using a standard drag law, added mass and lift forces, can not reproduce the turbulent suspension of particles in this configuration. In the following, the role of unresolved turbulent length scales smaller than the particle size is investigated.

\subsection{Evaluation of the finite-size correction model}

\begin{figure}
  \centerline{\includegraphics[width=0.6\textwidth]{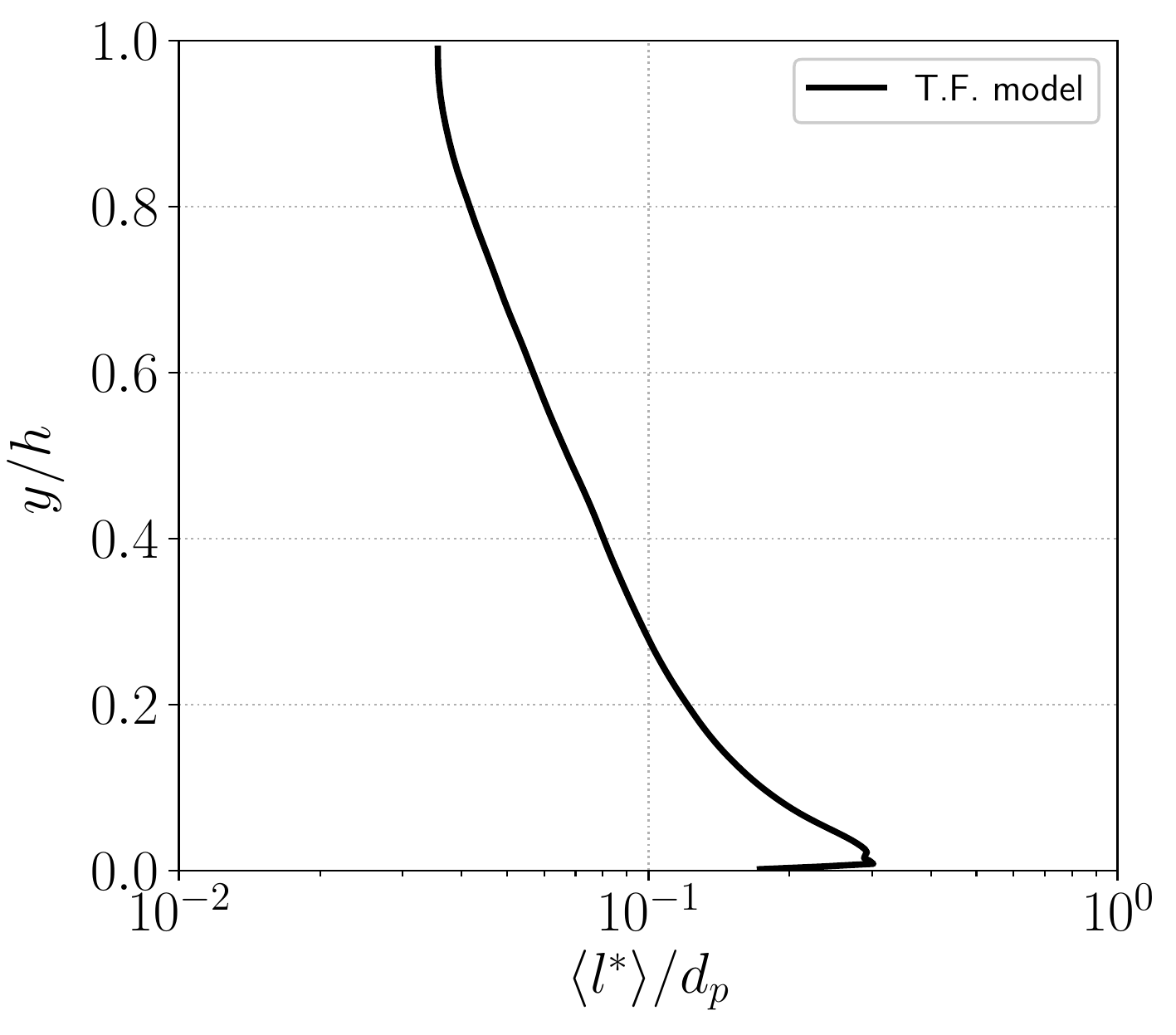}}
  \caption{Average profile of the length scale $l^*$ associated with the turbulent eddy having the same time scale as the particles as a fraction of $d_p$ from the two-phase simulation of configuration GB.}
\label{lstar}
\end{figure}

From the scaling analysis proposed by \cite{balachandar2009}, the relative velocity between the fluid phase and inertial particles is mainly influenced by an eddy having the same time scale as the particles with the corresponding length scale $l^*=\varepsilon^{f1/2}t_s^{3/2}$, with $\varepsilon^f$ the dissipation rate of fluid TKE. According to \cite{finn2016}, if $l^*>\Delta>d_p$, all the relevant flow scales are resolved and the particle dynamics can be accurately predicted. The average value of $l^*$ for configuration GB is calculated in the simulation and plotted in figure \ref{lstar}. It can be seen that $l^*/d_p<1$ and that $l^*$ decreases by one order of magnitude from the wall to the centreline of the channel. This result shows that, for this configuration, turbulent scales smaller than the particles can have a significant effect on the particle dynamics and may be responsible for the observed discrepancies.

\begin{figure}
  \centerline{\includegraphics[width=\textwidth]{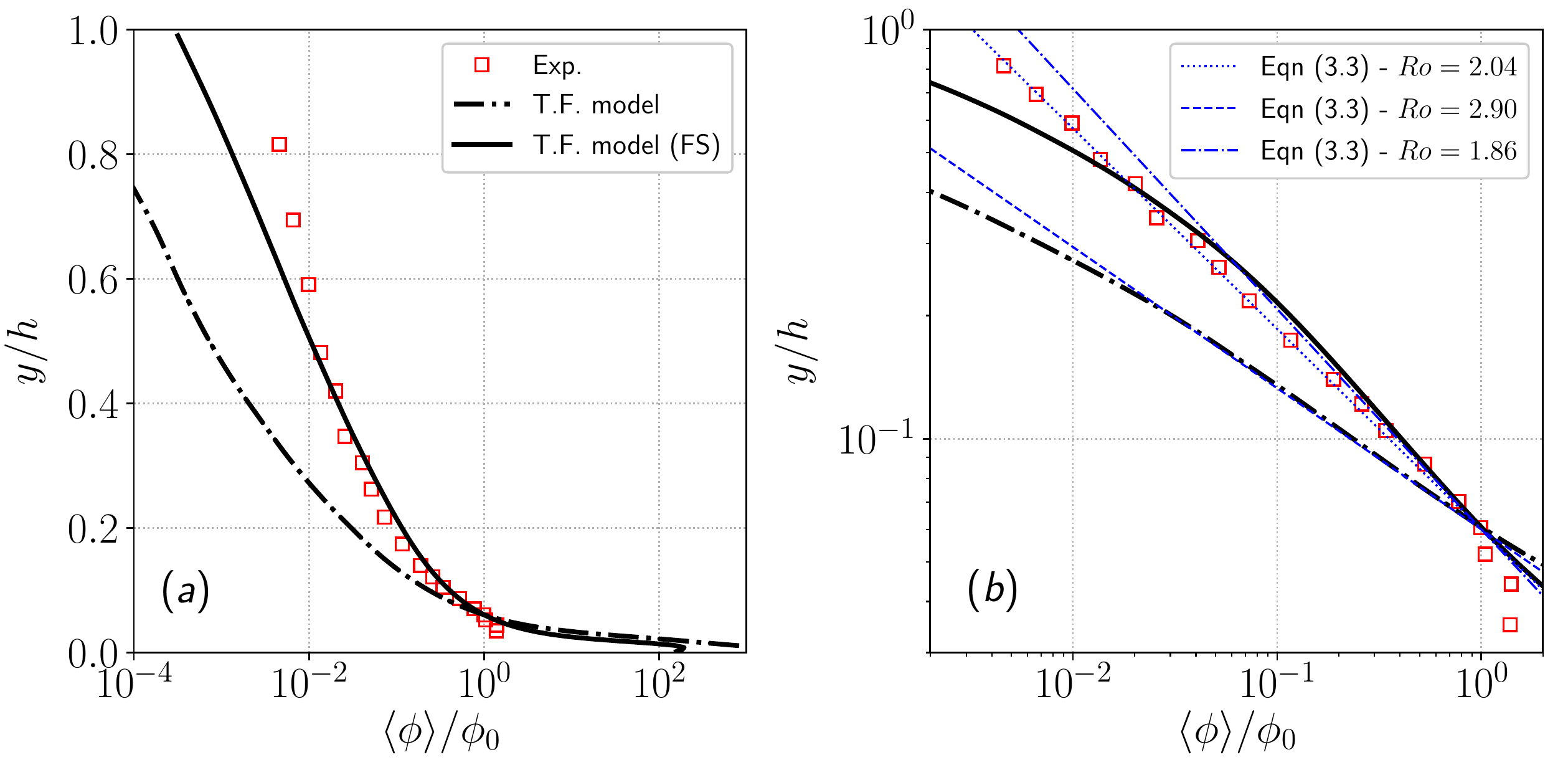}}
  \caption{Solid phase volumetric concentration profiles from the experiment (Exp.), two-phase flow simulation with finite-size correction model (T.F. model (FS)) and two-phase flow simulation without finite-size correction model (T.F. model) from configuration GB in semi-log scale in panel (a) and in log-log compared with analytical profiles from equation (\ref{rouse2}) with $Ro=2.04$, $Ro=2.90$ and $Ro=1.86$ in panel (b).}
\label{conc_fs}
\end{figure}

Given the broad range of length and time scales involved in a particle-laden horizontal boundary layer, multiple types of turbulence-particle interactions occur at different locations of the boundary layer. It is therefore crucial to develop a model applicable over a wide range of turbulence-particle interaction regimes. In the following, the finite-size correction model presented in \S\ref{fs_model} is tested for the three configurations GB, NS and NBS. The results of the simulations for the averaged concentration profile for configuration GB with and without the finite-size correction model are compared in figure \ref{conc_fs}. The prediction of the concentration profile by the two-phase flow model is significantly improved by the finite-size correction model without any tuning coefficient. The Rouse number predicted with the finite-size correction model is $Ro=1.86$ which is much closer to the experimental value compared with the prediction without correction (figure \ref{conc_fs}b). However, in the experiment the concentration profile is well described by the power law across the water depth whereas in the simulation the concentration decreases more rapidly toward the centreline of the channel.

\begin{figure}
  \centerline{\includegraphics[width=\textwidth]{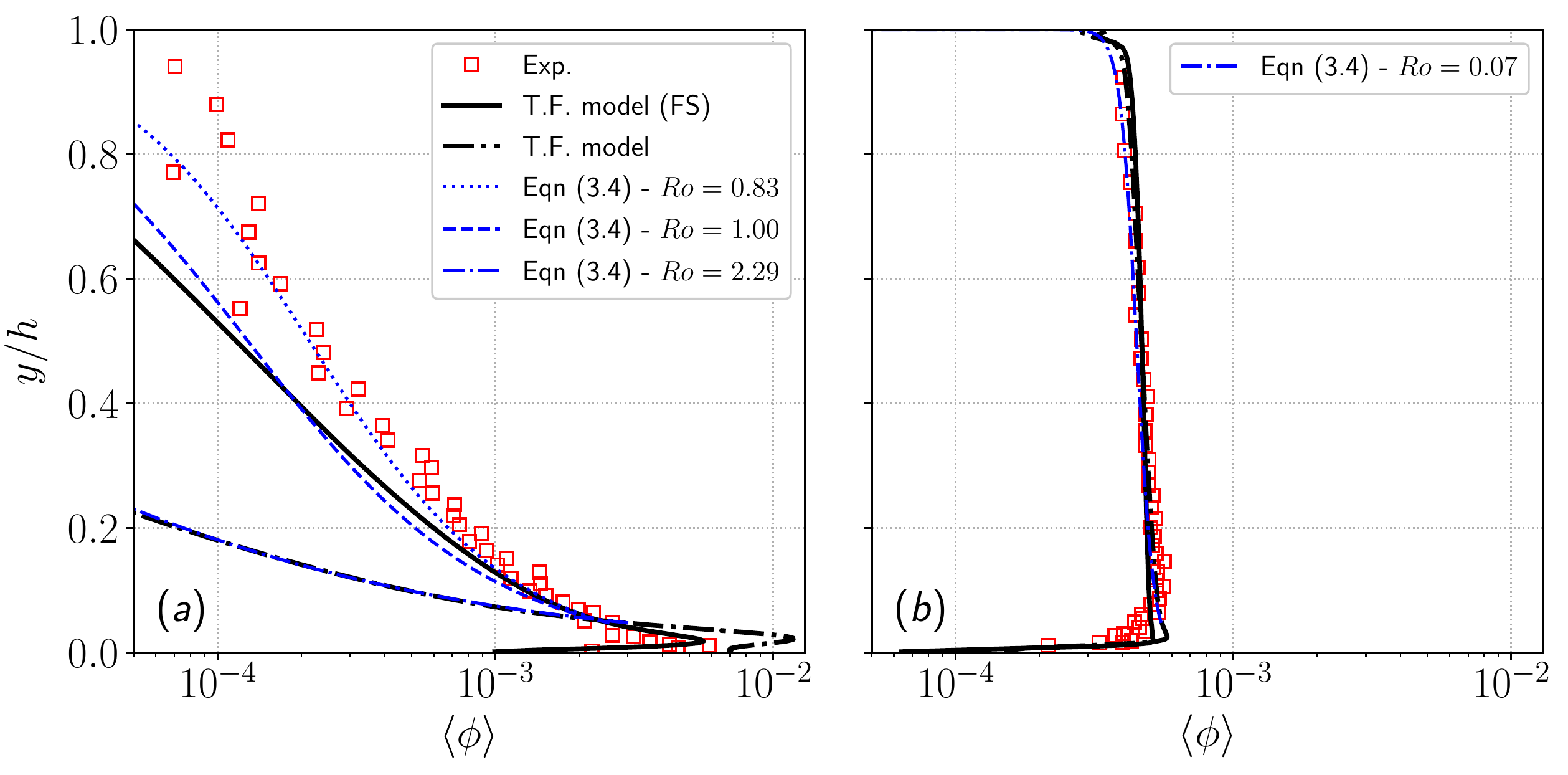}}
  \caption{Solid phase volumetric concentration profiles from the experiment (Exp.), two-phase flow simulation with finite-size correction model (T.F. model (FS)), two-phase flow simulation without finite-size correction model (T.F. model) and analytical profiles from equation (\ref{rouse2OC}) with $Ro=0.83$, $Ro=1.00$, $Ro=2.29$ and $Ro=0.07$ from configuration NS in panel (a) and configuration NBS in panel (b)}
\label{muste_conc}
\end{figure}

In order to further evaluate the finite-size correction model, the configurations NS and NBS are reproduced numerically using the two-phase flow model with finite-size correction. Analytical, experimental and numerical averaged concentration profiles are compared in figure \ref{muste_conc} for both configurations.

For configuration NS (figure \ref{muste_conc}a), the same conclusions as for configuration GB can be drawn. The finite-size correction model significantly improves the prediction of the turbulent suspension of particles without any tuning coefficient. The predicted Rouse number ($Ro=0.83$) is closer to the experimental value ($Ro=1.00$). The modelled concentration profile obtained using finite-size correction is in very good agreement with the experimental data compared with the simulation without the finite-size correction model.

For configuration NBS (figure \ref{muste_conc}b), the finite-size correction model does not alter the results predicted without finite-size correction. Almost no differences can be observed between the concentration profiles obtained with and without finite-size correction confirming that finite-size effects are negligible for configurations with a low suspension number. 

As a partial conclusion, it has been demonstrated that finite-size effects are important to predict turbulent suspension of inertial particles in a boundary layer flow when the suspension number is of the order of unity. The finite-size correction model proposed in this work significantly improves the model prediction for the average sediment concentration profile without the use of tuning parameter to fit the experimental data. 

\subsection{Lag velocity}

Another interesting feature of turbulent suspension of inertial particles is the existence of a velocity lag  between the average streamwise velocity of the fluid and of the particles \citep{kaftori1995b, nino1996, kiger2002, righetti2004, muste2005, kidanemariam2013}. \cite{kidanemariam2013}, based on fully resolved DNS, have been able to clearly identify the physical origin of this velocity lag as being due to the preferential concentration of suspended particles in low speed regions of the fluid flow which can be identified with ejection events. This velocity lag is not observed for particle-laden flows with a low suspension number \citep{muste2005} such as NBS but can be as high as 20\% of the bulk fluid velocity \citep{kidanemariam2013}. 
\begin{figure}
  \centerline{\includegraphics[width=\textwidth]{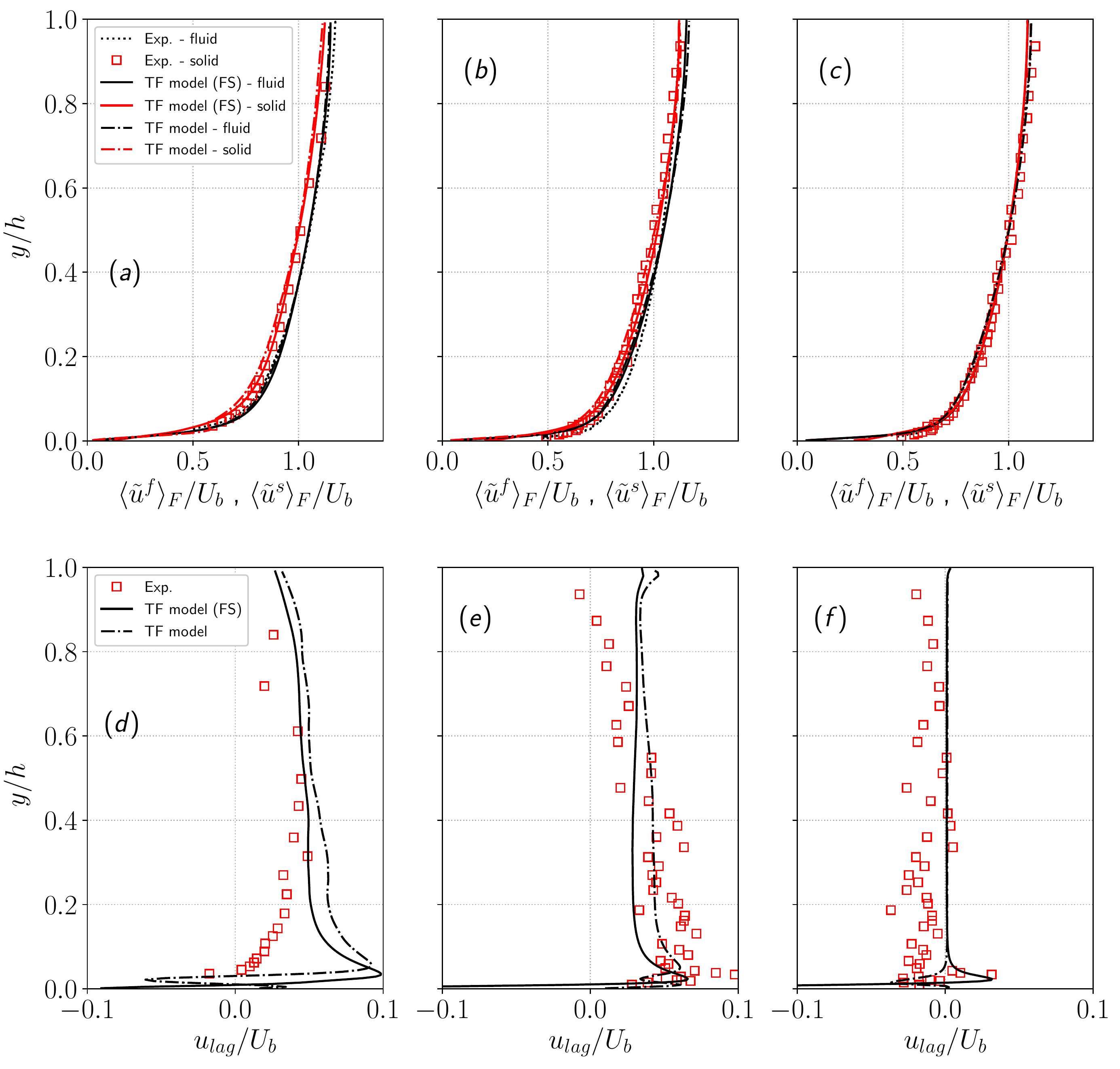}}
  \caption{Averaged fluid and solid velocity profiles in $(a, b, c)$ and lag velocity in $(d, e, f)$ from the two-fluid model with finite-size correction (T.F. model (FS)) and the two-fluid model without finite-size correction (T.F. model) for configurations GB (a, d), NS (b, e) and NBS (c, f) compared with experimental data (Exp.).}
\label{pl_vel}
\end{figure}

The averaged fluid and solid velocity profiles obtained numerically with and without the finite-size correction for configurations GB, NS and NBS are shown in the top panels of figure \ref{pl_vel}. The velocity profiles are in very good agreement with the experiments and they do not show much sensitivity to the finite-size correction model. The velocity difference is too small to be visible on these graphs, the lag velocity $u_{lag} = \langle \tilde u^f_i \rangle - \langle \tilde u^s_i \rangle$ is shown in the bottom panels of figure \ref{pl_vel}. For configurations GB and NS, the lag velocity is positive and of the order of 5-10\% of the bulk fluid velocity. The two-fluid model predicts the correct sign and order of magnitude for the lag velocity. The major discrepancy is observed in the near-wall region $y/h<0.2$ where the two-fluid model predicts a peak that is not observed in \cite{kiger2002} experiments. For the NS configuration, the lag velocity decreases linearly with the distance to the free surface. This is probably a free surface effect that is not fully captured by the symmetry plane boundary condition used in the present simulation, nevertheless, the model predictions are very satisfactory. Given the role played by the coherent structures of the flow in the velocity lag, the fact that the flow is over-resolved near the bottom boundary could explain the observed discrepancies for the GB case. However, since this difference is not observed for the other configurations, it might also be due to the high variability of the velocity measurements of \cite{kiger2002} near the bottom (up to 25\% for $y^+<50$). In both GB and NS configurations the finite-size correction model has a small influence on the lag velocity. In the NBS configuration the experimental data reveals a negligible lag velocity that even becomes negative. The two-fluid model with and without the finite-size correction model predicts a zero lag velocity except very near the bottom wall. From these three configurations, one can conclude that the existence of a lag velocity is not due to finite-size effects. More importantly, the fact that the model is able to recover the absence of lag velocity for NBS means that the two-fluid LES captures the physical mechanism correctly and can be used as a predictive tool to study this mechanism.


\subsection{Turbulent statistics}

Among the three configurations, the most accurate measurements of turbulent statistics have been obtained for configuration GB. In the following, this configuration is analysed in detail for the fluid and particle phase flow statistics. 

\begin{figure}
  \centerline{\includegraphics[width=\textwidth]{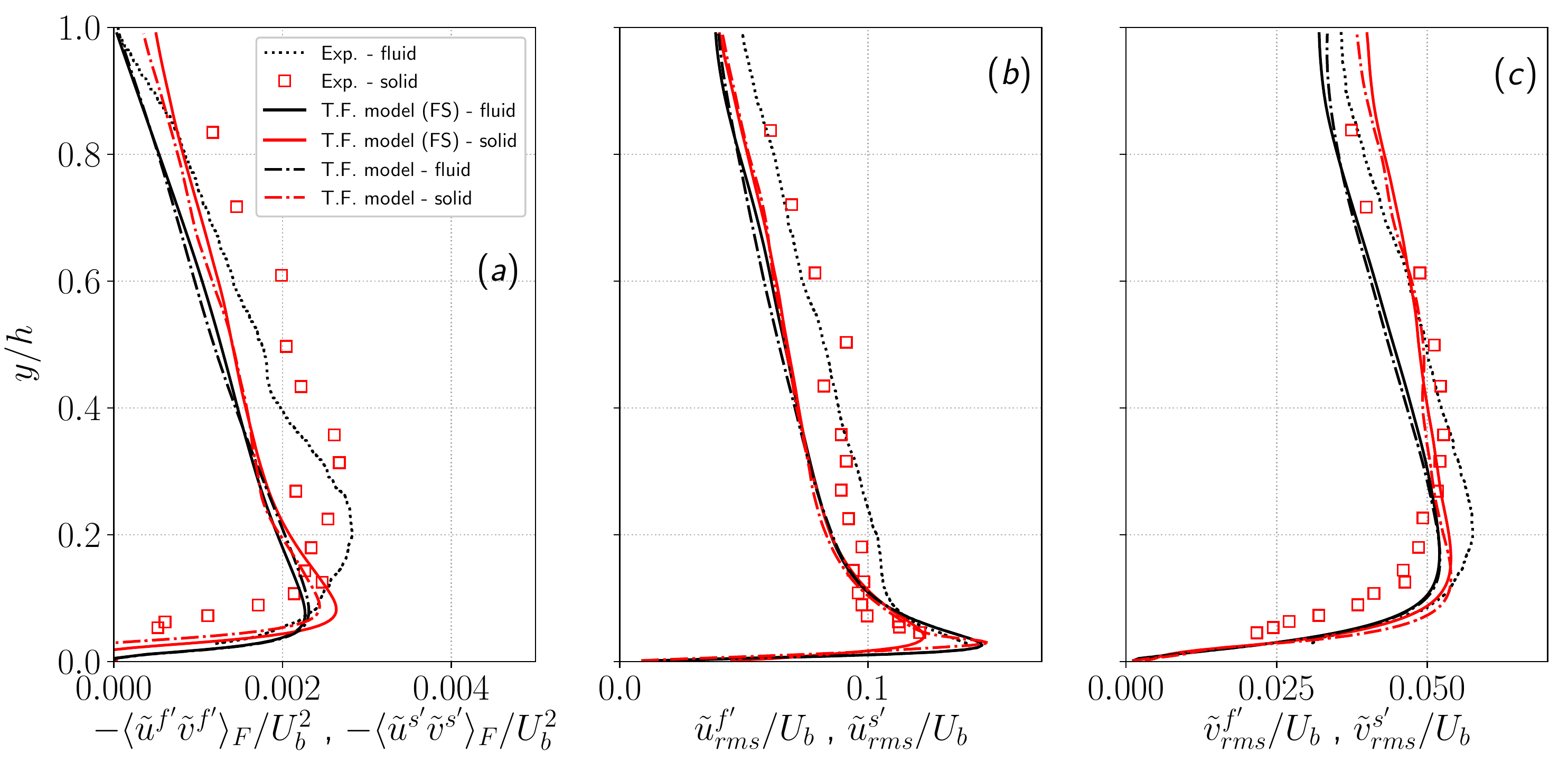}}
  \caption{Average profiles of fluid and solid Reynolds stress in $(a)$, r.m.s. of streamwise velocity fluctuations in $(b)$ and r.m.s of wall-normal velocity fluctuations in $(c)$ from the two-phase model with finite-size correction (T.F. model (FS)) and the two-phase model without finite-size correction (T.F. model) compared with experimental data (Exp.) from configuration GB.}
\label{pl_turb}
\end{figure}

 The wall-friction velocity for configuration GB predicted with and without finite-size correction is $u_\tau=2.70\times 10^{-2} \ m.s^{-1}$ and $u_\tau=2.72\times 10^{-2} \ m.s^{-1}$, respectively. Whereas the numerical wall-friction velocity is similar between the clear water and particle-laden configurations, the experiments suggest an increase of the friction velocity up to $u_\tau=2.99\times 10^{-2} \ m.s^{-1}$.  Averaged fluid and solid Reynolds stress and the r.m.s. of streamwise and wall-normal velocity fluctuation profiles from configuration GB with and without finite-size correction are compared with experimental data in figure \ref{pl_turb}. From figure \ref{pl_turb}a, the two-fluid model slightly under-estimates the fluid Reynolds shear stress compared with the experiments explaining the lower friction velocity in the simulations. However, experimental and numerical results are similar: the solid phase Reynolds shear stress is slightly greater than the fluid Reynolds shear stress away from the bottom wall. The maximum value for the solid Reynolds shear stress predicted by the two-fluid LES model is the same as in the experiments but the location is different. The r.m.s. of streamwise and wall-normal velocity fluctuations are in very good agreement with experimental results (figure \ref{pl_turb}b and \ref{pl_turb}c). As for the fluid Reynolds shear stress, the fluid phase velocity fluctuations are slightly under-estimated by the two-fluid model for $y/h>0.1$. For both experimental and numerical profiles, the r.m.s. of streamwise solid phase velocity fluctuations are equal near the centreline of the channel and becomes smaller in the near bottom wall region. Similarly to the observation of \cite{kidanemariam2013} in their fully resolved DNS, the two-fluid model predicts stronger wall-normal solid velocity fluctuations compared with the fluid away from the wall whereas experimental solid and fluid profiles are similar close to the channel centerline. The RMS of the solid velocity fluctuations decreases more rapidly than the fluid ones towards the wall. Overall, the turbulent statistics are not significantly affected by the finite-size correction model. 
 
The slight differences observed between the experimental and numerical fluid phase turbulent statistics come from the modulation of the turbulence by the particles. Again, from the scaling analysis by \cite{finn2016} and given the parameters of the configuration from \cite{kiger2002}, the presence of the particles is expected to dampen fluid turbulence, whereas in the experiments a slight increase of the Reynolds stress and velocity fluctuations are observed compared with the clear water configuration. Indeed, according to \cite{balachandar2009}, the turbulence enhancement due to the presence of the particles comes from the combined action of the oscillating wakes behind particles having a high particle Reynolds number. The conjugate action of all the wakes of the particles participates to increase the overall fluid turbulence.

To be able to predict the turbulence enhancement, the two-fluid model should have the capacity to capture the vortex shedding behind the particles by fully resolving the fluid/solid interface which is not the case for the Eulerian-Eulerian two-phase flow model. Nevertheless, the turbulence enhancement due to the particles is not a dominant mechanism in this configuration. According to \cite{finn2016}, the net production of turbulence by the particles is dominant for particle Reynolds numbers higher than Reynolds number $\Rey_p=400$ even if oscillatory wakes behind particles can be observed for lower $\Rey_p$ depending on flow properties, particle shape or distance from the wall for example. In the present configuration, the particle Reynolds number based on the scaling relations from \cite{finn2016} is equal to $\Rey_{p} = 4.8$ and the maximum particle Reynolds number predicted in the simulation is $\Rey_{p,max}\approx20$, which is significantly below the threshold Reynolds number of 400.

Flow hydrodynamics and turbulent statistics are in good agreement with experimental data and the overall relative behaviour between the fluid and solid phase is correctly captured by the two-phase flow model. The fact that the two-fluid flow model does not resolve the particle-fluid interface implies that the turbulence enhancement induced by the presence of the particles is not resolved. However, for such flow and particle parameters, according to the scaling analyses from \cite{balachandar2009} and \cite{finn2016}, this mechanism is not dominant. The lower fluid velocity fluctuations predicted by the two-fluid model near the channel centreline only results in a slight under-estimation of the sediment concentration in the same region compared with the experiments. 

\begin{figure}
  \centerline{\includegraphics[width=\textwidth]{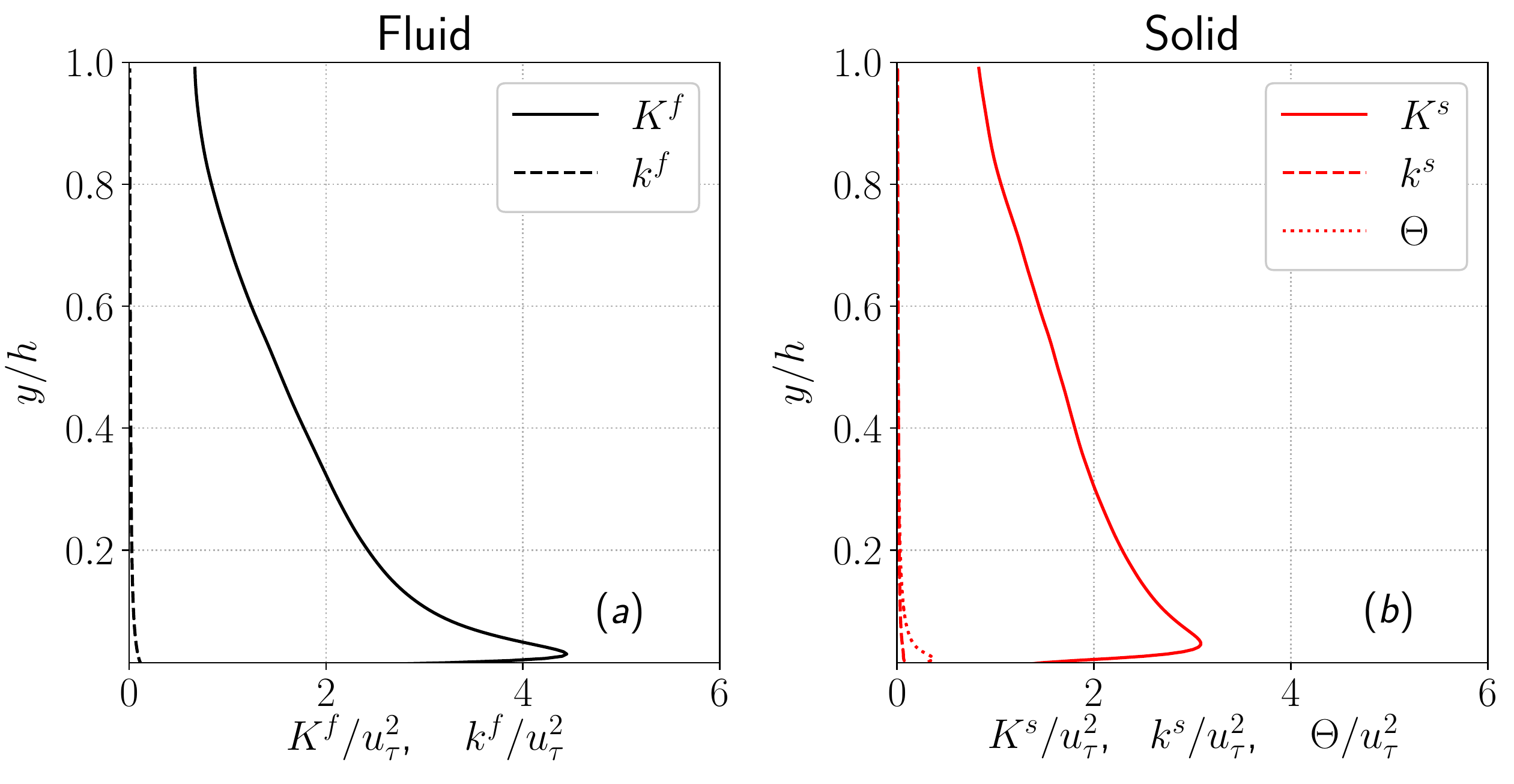}}
  \caption{Resolved and sub-grid fluid phase TKE $K^f$ and $k^f$ in (a), resolved and sub-grid solid phase TKE $K^s$ and $k^s$ and granular temperature $\Theta$ in (b) made dimensionless by the friction velocity $u_\tau$ from configuration GB.}
\label{pl_tke}
\end{figure}

From the resolved and sub-grid TKE profiles for the fluid and solid phases presented in figure \ref{pl_tke}, it appears that most of the TKE is resolved ($k^f/K^f < 5\%$ and $k^s/K^s< 5\%$). The fact that the solid phase sub-grid TKE is equal to zero through the channel height shows that the resolution is very close to DNS and validates the hypothesis made in \S\ref{tpmodel_eq} to neglect the sub-grid terms. Furthermore, $K^s \gg \Theta$ in the channel except very near the solid boundary ($y/h<0.05$) showing that kinetic and collisional dispersive forces should not be dominant compared with the drag force to suspend the solid particles for such dilute configurations. This hypothesis is further investigated in \S\ref{discussion}.

\subsection{Volume fraction sensitivity}

An additional simulation (configuration NS*) is performed to evaluate the robustness of the proposed model to an increase of the mass loading. Experimental data from \cite{muste2005} using NS having higher volume fractions compared with the previous NS configuration is reproduced numerically using the two-fluid model. The hydrodynamic and particle parameters are the same as for the NS configuration but the total solid phase volume fraction is multiplied by a factor 3.5 ($\phi_{tot}=16.2\times10^{-4}$). The concentration can still be considered dilute and particles do not form a settled bed at the bottom of the channel.

\begin{figure}
  \centerline{\includegraphics[width=0.6\textwidth]{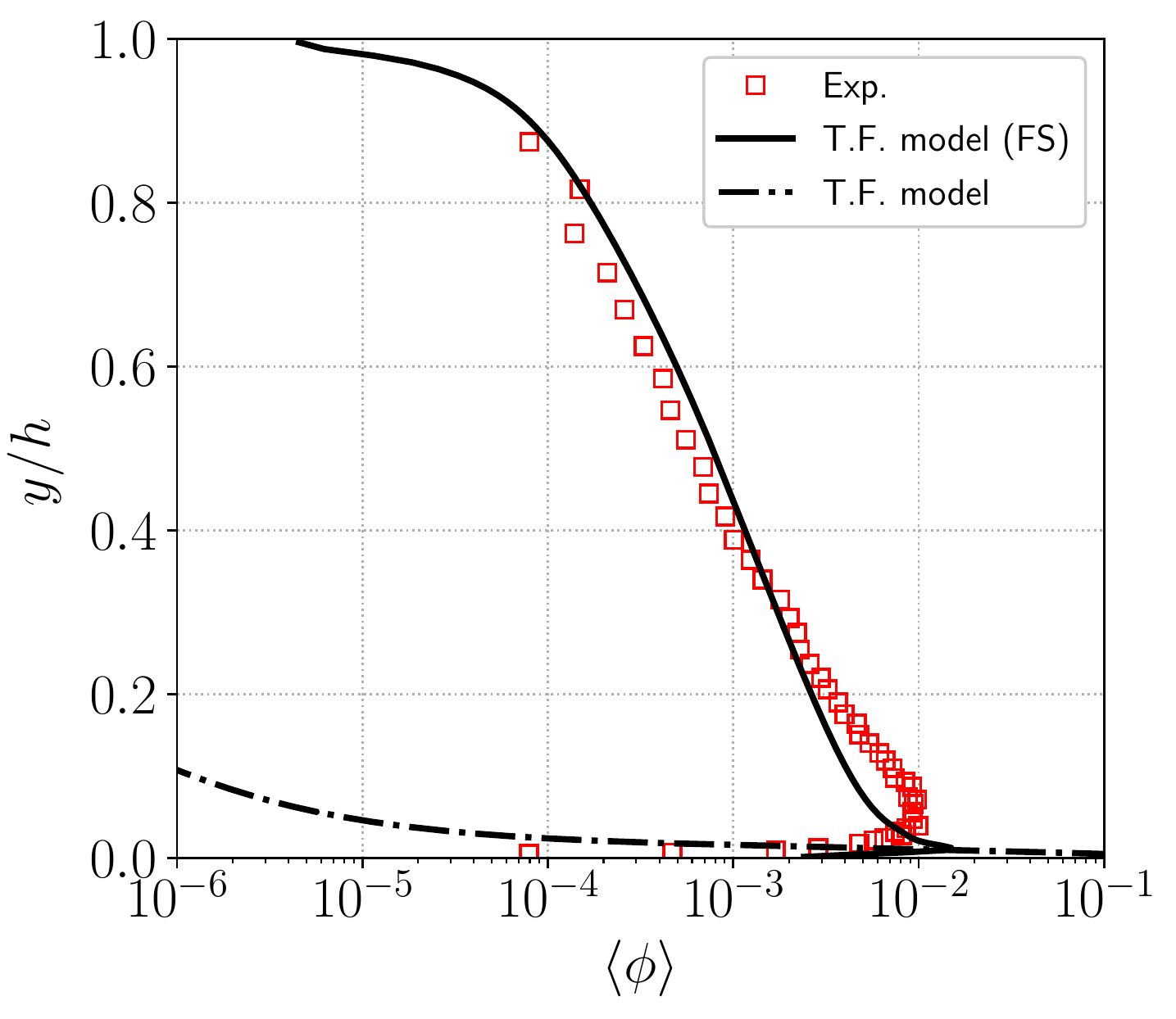}}
  \caption{Solid phase volumetric concentration profiles from the experiment (Exp.), two-phase flow simulation with finite-size correction model (T.F. model (FS)), two-phase flow simulation without finite-size correction model (T.F. model) from configuration NS*.}
\label{conc_sens}
\end{figure}

The averaged concentration profile from the simulation of configuration NS with higher volume fractions is compared with experimental data in figure \ref{conc_sens}. As observed in the previous sections, the agreement with experimental data is significantly improved by the finite-size correction for natural sediments. This result is even more spectacular considering that without the correction model, the particles settle almost completely at the bottom of the channel resulting in an even larger under-estimation of the suspension of particles at higher volume fractions by the original two-fluid model. 
%
%
\section{Discussion}
\label{discussion}
In this section the sensitivity of the model to the different components of the finite-size correction model are discussed as well as the sensitivity to the grid/second filter resolution is presented. 

\subsection{Relative influence of the different terms of the finite-size correction model}

\begin{figure}
  \centerline{\includegraphics[width=0.9\textwidth]{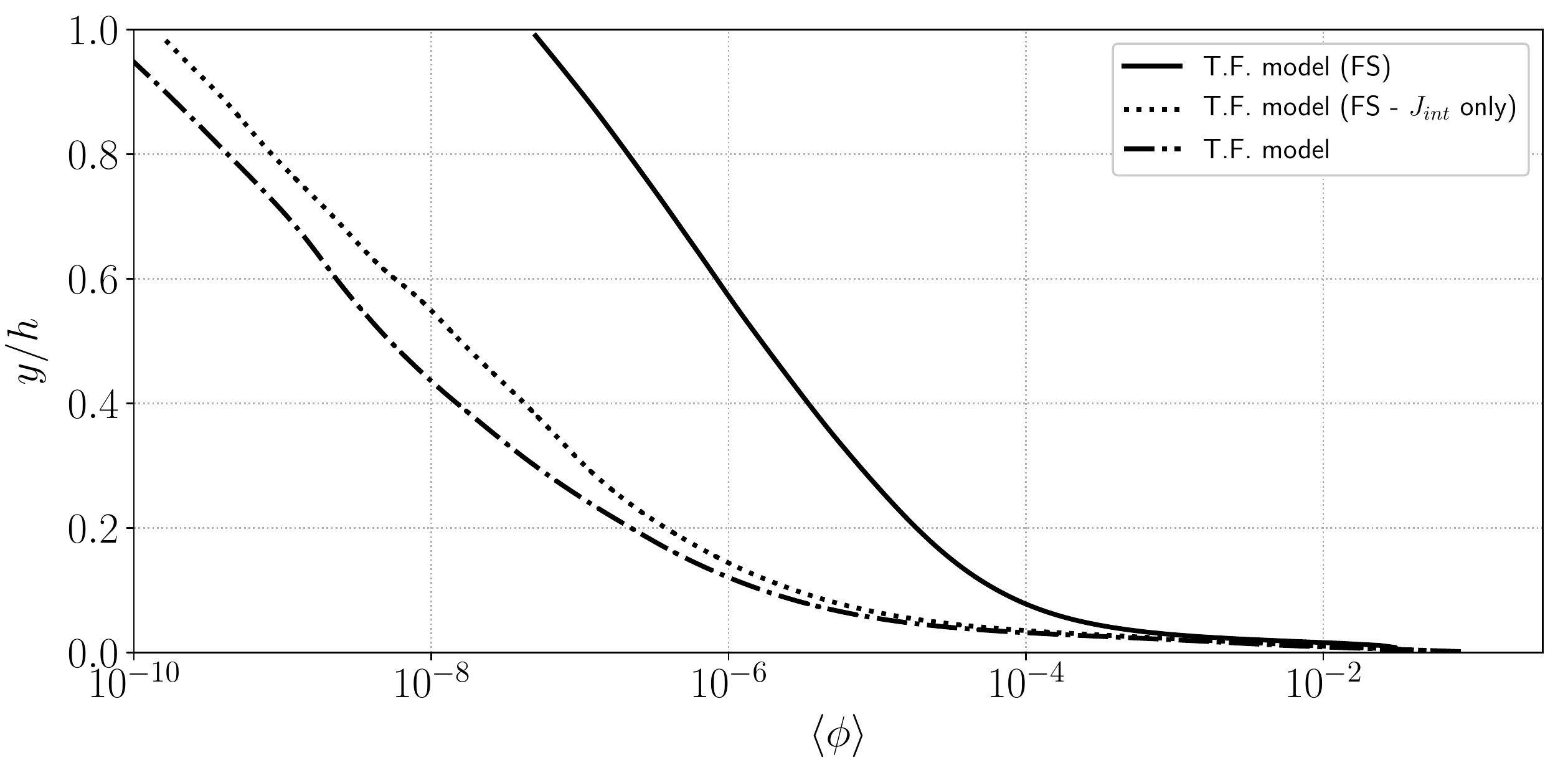}}
  \caption{Solid phase volumetric concentration profiles from the two-phase flow simulation with finite-size correction model (T.F. model (FS)), two-phase flow simulation with finite-size correction only in the production term of granular temperature (T.F. model (FS-$J_{int}$ only) and two-phase flow simulation without finite-size correction model (T.F. model) from configuration GB in semi-log scale.}
\label{relcontrib}
\end{figure}

In order to evaluate the relative influence of the modified drag law and the modified production of granular temperature, a new simulation is performed for which finite-size effects are taken into account only in the production term of the granular temperature transport equation. In other words, the simulation is performed using the drag law from equation (\ref{drag}) and the production of granular temperature from equation (\ref{jintfs}). The average concentration profile obtained from the simulation including finite-size effects only in the production of granular temperature equation is compared with the concentration profile from the simulations with and without finite-size correction in figure \ref{relcontrib}.

The concentration profile obtained from the two-fluid simulation including finite-size effects only in the production term of the granular temperature is similar to the profile without the finite-size correction model. Indeed, for dilute particles, fluid-particle interactions are dominant compared with particle-particle interactions. The slope of the concentration profile in dilute regions of the flow is shaped by the drag force and the modification of the production term of granular temperature has almost no effect. However, the effect of the modification of the granular temperature transport equation could become dominant for higher concentrations. It should be noted that the modification of granular temperature transport equation is necessary because a simulation including finite-size effects only in the drag law and not in the production term of granular temperature was shown to be highly unstable. The observed instability is not the result of a numerical issue but rather the consequence of a physical inconsistency in the energy transfers between the two phases. Indeed, including finite-size effects only in the drag law and not in the production term of granular temperature is not physically accurate. Some of the turbulent flow scales are not taken into account in the energy budget between the phases making the simulation unstable.

\begin{figure}
  \centerline{\includegraphics[width=0.6\textwidth]{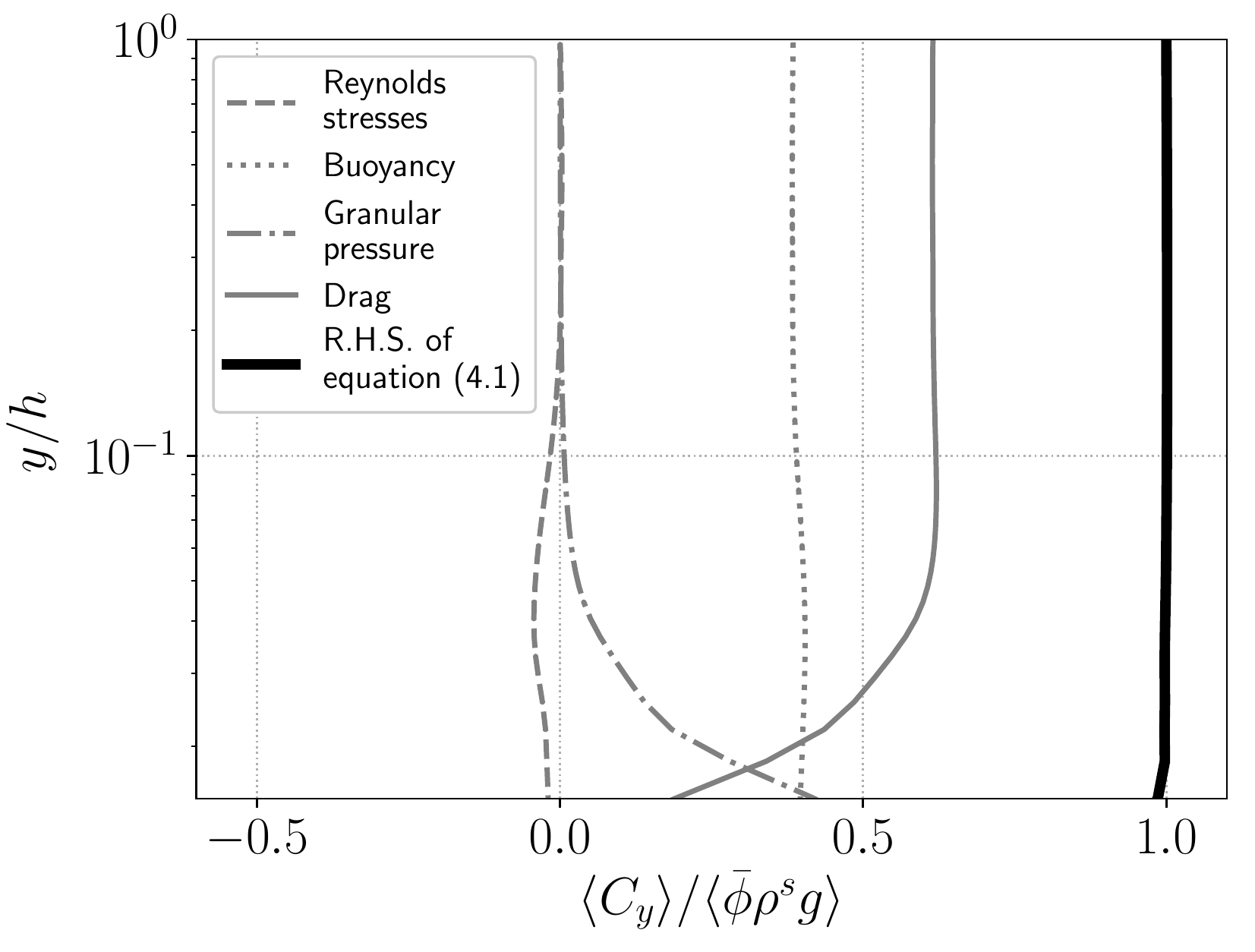}}
  \caption{Averaged contributions to the wall-normal momentum budget $\langle C_y \rangle$ on the right-hand side (R.H.S.) of the momentum balance (\ref{mom_balance}) and their sum as a fraction of the gravity force $\langle\bar\phi\rho^sg\rangle$  from configuration GB.}
\label{mom_budg}
\end{figure}

The averaged wall-normal momentum balance for the solid phase can be written as
\begin{equation}
\label{mom_balance}
\underbrace{\bigg\langle\bar\phi \rho^s g\bigg\rangle}_{\parbox{2cm}{\centering Gravity}}  = 
\bigg\langle\underbrace{\frac{\partial R^s_{xy}}{\partial x} + \frac{\partial R^s_{yy}}{\partial y} + \frac{\partial R^s_{zy}}{\partial z}}_{\parbox{2cm}{\centering Reynolds \\ stresses}} \bigg\rangle
+ \underbrace{\bigg\langle\bar\phi\frac{\partial \bar P^f}{\partial y}\bigg\rangle }_{\parbox{2cm}{\centering Fluid pressure \\ (Buoyancy)}}
+ \underbrace{\bigg\langle\frac{\partial \bar P^s}{\partial y}\bigg\rangle }_{\parbox{2cm}{\centering Granular \\ pressure}}
+  \underbrace{\bigg\langle\bar I_y\bigg\rangle}_{\parbox{2cm}{\centering Drag}}
\end{equation}
with $R_{ij}=\partial\langle \rho^s \bar\phi \tilde u_i^{s'}\tilde u_j^{s'}\rangle/\partial x_j$ the Reynolds shear stress coming from averaging of the nonlinear advection terms with $\tilde u_i^{s'}= \tilde u_i^s - \langle \tilde u_i^s \rangle$ the solid phase resolved velocity fluctuations. In figure \ref{mom_budg} the four terms of the right-and side of equation \ref{mom_balance} are plotted in dimensionless form, normalized by $\langle\rho^s\bar\phi g\rangle$, in semi-log scale for $y/h$. This figure shows that the predicted suspended particle concentration profile results from a balance between gravity, buoyancy and drag forces in the upper part of the channel. In such dilute systems the effect of dispersive kinetic and collisional forces are not significant except very near the wall $y/h<0.05$ ($y/d_p<5$). This supports the hypothesis that the discrepancies observed in the original model can not be due to a flaw in the kinetic theory formulation but are due to fluid-particle interaction forces.  

\subsection{Second filter size sensitivity}
\label{sens_test}
As mentioned in \S\ref{fs_model}, the width of the second filter $\breve\Delta$ should not be too small to be free from disturbances generated by the presence of the particles. On the other hand, the second filter size should not be too large in order to provide an accurate representation of the velocity ``seen'' by the particles. The minimum filter width $\breve\Delta_{min}=2d_p$ has been determined from \cite{kidanemariam2013}, but there is no clear criteria for the maximum filter width. However, for computational efficiency, since the second filter width depends on the spatial discretization in the streamwise and spanwise directions for this configuration, it can be crucial to determine the maximum acceptable filter width to accurately predict the average concentration profile with coarser grid resolutions.

\begin{table}
  \begin{center}
\def~{\hphantom{0}}
  \begin{tabular}{lcccc}
     Mesh &  ~~$N_x\times N_y \times N_z$ ~ & ~$\Delta_x^+$, $\Delta_z^+$~ & ~$\Delta_y^+$ (bottom)~ & $\breve\Delta$\\[3pt]
     M1 &  ~~$314\times220\times160$~~ & $11$ & 1 & $2d_p$\\
     M2 &  ~~$210\times147\times107$~~ & $17$ & 1.5 & $3d_p$\\
     M3 &  ~~$126\times88\times63$~~ & $22$ & 2 & $4d_p$\\
     M4 &  ~~$80\times56\times40$~~ & $44$ & 4 & $8d_p$\\
  \end{tabular}
  \caption{Mesh characteristics for the second filter size sensitivity test.}
  \label{meshes}
  \end{center}
\end{table}

\begin{figure}
  \centerline{\includegraphics[width=\textwidth]{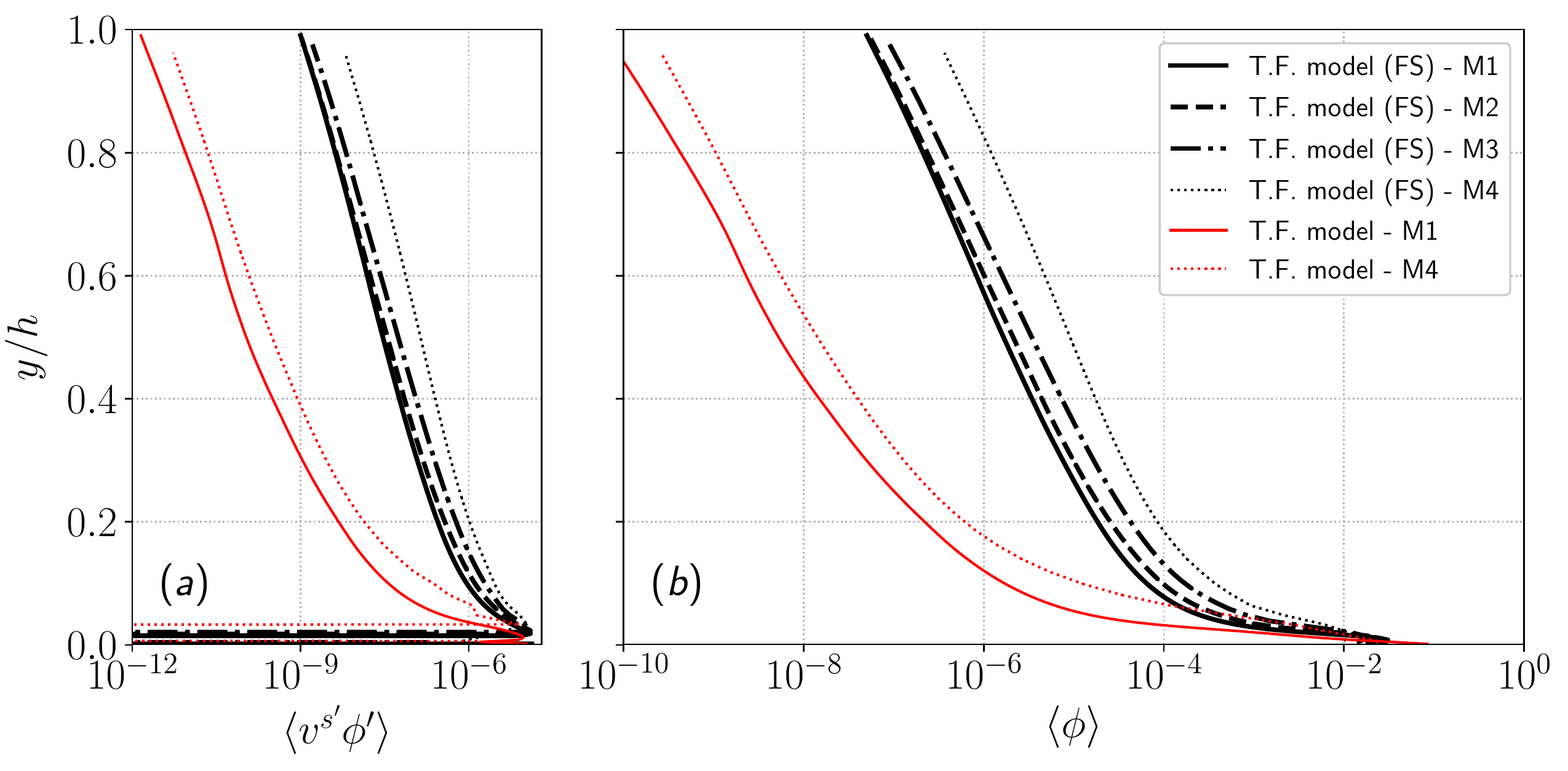}}
  \caption{Averaged Reynolds flux (panel (a)) and solid phase volumetric concentration (panel(b)) profiles from two-phase flow simulations with finite-size correction model (T.F. model (FS)) from configuration GB using mesh M1, M2, M3 and M4 and without finite-size correction model (T.F. model) using mesh M1 and M4.}
\label{kigerandpan_meshsens}
\end{figure}

Additional simulations for configuration GB with different mesh resolutions have been performed to measure the influence of the spatial discretization and the second filter width $\breve\Delta$ on the sediment concentration profile prediction. The mesh characteristics for the different simulations are presented in table \ref{meshes}. The comparison between Reynolds fluxes and concentration profiles with and without finite-size correction obtained with mesh M1, M2, M3 and M4 are presented in figure \ref{kigerandpan_meshsens}. 


The turbulent dispersion of the particles increases for a coarser resolution (figure \ref{kigerandpan_meshsens}a). As a consequence, the amount of suspended particles in the water column predicted by the two-fluid model increases with increasing filter width (figure \ref{kigerandpan_meshsens}b). The difference between the concentration profiles from simulations using the finite-size correction model with $\breve\Delta=2 d_p$ (mesh M1) and $\breve\Delta=3 d_p$ (mesh M2) is negligible and the agreement can still be considered as acceptable for a filter width of $\breve\Delta=4 d_p$ (mesh M3). However, discrepancies become important for a larger filter width ($\breve\Delta=8 d_p$ with mesh M4). More quantitatively, the concentration profile converges at first order with the reference simulation results from mesh M1 for increasing vertical resolution. 

Even without the finite-size correction model, the Reynolds flux is increased between simulations using mesh M1 and mesh M4 (figure \ref{kigerandpan_meshsens}a) suggesting that the over-prediction of the concentration does come from the finite-size correction model only but also from the modification of the flow hydrodynamic for coarser grid resolutions.  

As a conclusion, in order to accurately predict the concentration profile, the sensitivity analysis suggests that the grid resolution at the wall should not exceed 4 wall units $\Delta_y^+ < 4$. It is mandatory to have at least one grid point in the laminar sub-layer in order to resolve the turbulent coherent flow structures in the near-wall region and to use a second filter smaller than $8 d_p$  ($\breve\Delta < 8d_p$) to accurately resolve the fluid velocity ``seen'' by the particles.

\section{Conclusion}

Turbulence-particle interactions may play a key role in particle-laden flows by modifying the turbulent dispersion of particles by turbulent eddies and by the feedback of particles on the turbulent eddies. From a modelling point of view, a specific challenge is the huge range of cascading turbulent eddy sizes $O(10^{-1}-10^{-4})$ m and their interactions with different grain sizes $O(10^{-3}-10^{-5})$ m. The very wide range of length scales involved does not allow us to systematically use the turbulence-resolving approach at the particle scale to address this problem due to its prohibitive computational cost and turbulence-resolving continuum approaches, such as the two-fluid LES approach, are needed.

In this paper, the two-fluid LES method has been tested against experimental data and a finite-size correction model has been developed. The new model has been validated against available experimental data for the dilute turbulent suspension of finite-sized particles transported by a boundary layer flow. The improved model has been shown to  accurately predict the suspended particle concentration profile as well as the existence of a streamwise lag velocity for heavier-than-fluid particles without the use of any tuning parameter to fit the experimental data. In the proposed correction model, a distinction is made between turbulent flow scales larger or smaller than the particle diameter. The velocity field ``seen'' by the particles in the drag law is filtered at a scale $\breve\Delta \geq 2d_p$ and smaller turbulent scales contribute to reduce the particle response time by the addition of a sub-particle scale eddy viscosity to the molecular viscosity in the particle Reynolds number definition. The second effect of the correction is to increase the production of granular temperature by a modification of the source term in the granular temperature equation. While modification of the drag law is more important for the accurate prediction of the suspended particle concentration profile in the dilute configuration investigated herein, the modification of the granular temperature equation is mandatory for the physical consistency and the numerical stability of the model. At last, the sensitivity analysis of the model results for the second filter size $\breve\Delta$ has shown that the grid resolution could be as high as 4 particle diameters without loss of accuracy as long as one grid point is located in the laminar sub-layer.

The work presented herein is an important step towards two-fluid LES of more complex applications such as scour around hydraulic structures, wave-driven sediment transport or turbidity currents to cite a few geophysical flow examples. In all the aforementioned applications, additional complex interaction mechanisms such as interactions with a sediment bed occur. The two-fluid approach should therefore be validated against configurations involving higher sediment volume fractions compared with the present configurations.

Furthermore, detailed measurements of turbulence-particle interactions are really challenging and using turbulence-resolving simulations in addition to measurements is probably the only way to improve our understanding of the role of these mechanisms on particle transport dynamics. \\

\noindent{\bf Acknowledgments}. Julien Chauchat and Antoine Mathieu would like to thank R\'emi Zamansky for the fruitful discussions about the finite-size correction model. The authors would like to acknowledge the financial support from Agence de l'Innovation de D\'efense (AID), Shom through project MEPELS and Agence Nationale de la Recherche (ANR) through project SheetFlow (ANR-18-CE01-0003). Tian-Jian Hsu and Julien Chauchat also like to acknowledge support from the Munitions Response Program of the Strategic Environmental Research and Development Program under Project MR20-S1-1478. Most of the computations presented in this paper were performed using the GENCI infrastructure under Allocations A0060107567 and A0080107567 and the GRICAD infrastructure. \\

\noindent{\bf Declaration of Interests}. The authors report no conflict of interest. \\

\appendix
\section{Dynamic Lagrangian model from \cite{meneveau1996}}
\label{dynlag}

As a result of the nonlinear advection terms filtering, additional sub-grid terms need to be modelled $\sigma^{a,sgs}_{ij}=\rho^a\bar\phi^a(\widetilde{u^a_iu^a_j} - \tilde u^a_i \tilde u^a_j)$, with subscript $a=\{f,s\}$ denoting the fluid or solid phase, $\rho^a$ and $\bar\phi^a$ the density and filtered volumetric concentration of phase $a$, respectively. The most common way to model the sub-grid stress is to use the Smagorinsky model,
\begin{equation}
 \sigma^{a,sgs}_{ij} = 2\rho^a\overline\phi^a\Delta^2 \vert \tilde{\boldsymbol S}^a \vert \left(C_1^a\tilde S_{ij}^a - \frac{1}{3}C_2^a\tilde S_{kk}^a\delta_{ij}\right),
\end{equation}
with $\Delta$ the filtered width, $\tilde{\boldsymbol S}^a$ resolved strain rate tensor of phase $a$ and $C_1^a$ and $C_2^a$ the model coefficients. To adjust the model coefficients, a dynamic procedure samples the turbulent stress from the smallest resolved scales and extrapolate to determine the turbulent stress associated with unresolved turbulent scales below $\Delta$.

The starting point to determine the first coefficient $C_a^a$ is the algebraic identity
\begin{equation}
\label{identity}
  \mathcal{L}_{ij}^a = \mathcal{T}_{ij}^a - \mathcal{\tau}_{ij}^a,
\end{equation}
relating the turbulent stress associated to two different filter widths $\Delta$ and $\hat\Delta=2\Delta$ with

\begin{equation}
\mathcal{L}_{ij}^a=\widehat{\tilde u^a_i \tilde u^a_j} - \widehat{\tilde u}^a_i \widehat{\tilde u}^a_j, \quad\quad \mathcal{T}_{ij}^a=\widehat{\widetilde{u^a_iu^a_j}} - \widehat{\tilde u}^a_i \widehat{\tilde u}^a_j \quad \mbox{and\ } \quad \mathcal{\tau}_{ij}^a=\widehat{\widetilde{u^a_iu^a_j}} - \widehat{\tilde u^a_i \tilde u^a_j}.
\end{equation}

The Smagorinsky model is used to model the turbulent stress $\mathcal{\tau}_{ij}^a$ at scale $\Delta$ and $\mathcal{T}_{ij}^a$ at scale $2\Delta$ following:

\begin{equation}
\label{delta}
  \mathcal{\tau}_{ij}^a = -2C_1^a\Delta^2\vert \tilde{\boldsymbol S}^a \vert{\tilde S_{ij}^a},
\end{equation}

\begin{equation}
\label{deltahat}
  \mathcal{T}_{ij}^a = -2C_1^a(2\Delta)^2\vert \widehat{\tilde{\boldsymbol S}^a} \vert\widehat{\tilde S_{ij}^a}.
\end{equation}

Replacing expressions (\ref{delta}) and (\ref{deltahat}) in the identity (\ref{identity}) and minimizing the mean square error between the resolved identity and the Smagorinsky model leads to the expression for the coefficient $C_1^a$,
\begin{equation}
  C_1^a=\frac{\mathcal{F}_{LM}^a}{\mathcal{F}_{MM}^a},
\end{equation}
with $\mathcal{F}_{LM}^a$ and $\mathcal{F}_{MM}^a$ the products $\mathcal{L}_{ij}^a\mathcal{M}_{ij}^a$ and $\mathcal{M}_{ij}^a\mathcal{M}_{ij}^a$ with $\mathcal{M}_{ij}^a=2\Delta^2\left[\widehat{\vert\tilde S^a\vert\tilde S_{ij}^a}-4\vert\widehat{\tilde S}^a\vert \widehat{\tilde S}_{ij}^a\right]$ averaged over streamlines following the expressions:
\begin{equation}
\label{int1}
  \mathcal{F}_{LM}^a=\int^t_{-\infty}\mathcal{L}_{ij}^a\mathcal{M}_{ij}^a(z(t'), t')W(t-t')dt',
\end{equation}
\begin{equation}
\label{int2}
  \mathcal{F}_{MM}^a=\int^t_{-\infty}\mathcal{M}_{ij}^a\mathcal{M}_{ij}^a(z(t'), t')W(t-t')dt'.
\end{equation}
The the weighting function $W(t-t')$ is used to control the relative importance of the events near time $t$ with those of earlier times. The exponential form $W(t-t')=T^{-1}e^{-(t-t')/T}$, with $T=\frac{3\Delta}{2}\left(\mathcal{F}_{LM}^a\mathcal{F}_{MM}^a\right)^{-\frac{1}{8}}$, makes the integrals (\ref{int1}) and (\ref{int2}) solutions of the transport equations
\begin{equation}
  \frac{\partial \overline\phi^a \mathcal{F}_{LM}^a}{\partial t} + \frac{\partial\overline\phi^a \tilde u_j^a\mathcal{F}_{LM}^a}{\partial x_i} = \frac{2}{3\Delta}(\mathcal{F}_{LM}^a\mathcal{F}_{MM}^a)^{\frac{1}{8}}(\mathcal{L}_{ij}^a\mathcal{M}_{ij}^a - \mathcal{F}_{LM}^a),
\end{equation}
and
\begin{equation}
  \frac{\partial \rho^a \overline\phi^a  \mathcal{F}_{MM}^a}{\partial t} + \frac{\partial \rho^a \overline\phi^a \tilde u_j^a\mathcal{F}_{MM}^a}{\partial x_i} = \frac{2}{3\Delta}(\mathcal{F}_{LM}^a\mathcal{F}_{MM}^a)^{\frac{1}{8}}(\mathcal{M}_{ij}^a\mathcal{M}_{ij}^a - \mathcal{F}_{MM}^a).
\end{equation}

To compute the second model coefficient $C_2^a$, a similar procedure is used. Using the following identity for the spherical part of the sub-grid scale shear stress tensor:
\begin{equation}
  \mathcal{L}^{*,a} =\mathcal{T}^{*,a} - \mathcal{\tau}^{*,a}.
\end{equation}
Here
\begin{equation}
\mathcal{L}^{*,a}=\frac{1}{3}tr(\widehat{\tilde u^a_i \tilde u^a_j} - \widehat{\tilde u}^a_i \widehat{\tilde u}^a_j), \quad\quad \mathcal{T}^{*,a}=\frac{1}{3}tr(\widehat{\widetilde{u^a_iu^a_j}} - \widehat{\tilde u}^a_i \widehat{\tilde u}^a_j) \quad \mbox{and\ } \quad \mathcal{\tau}^{*,a}=\frac{1}{3}tr(\widehat{\widetilde{u^a_iu^a_j}} - \widehat{\tilde u^a_i \tilde u^a_j}),
\end{equation}
modeled following
\begin{equation}
  \mathcal{\tau}^{*,a} = -\frac{2}{3}C_2^a\Delta^2\vert \tilde{\boldsymbol S}^a \vert{\tilde S_{kk}^a\delta_{ij}},
\end{equation}

\begin{equation}
  \mathcal{T}^{*,a} = -\frac{2}{3}C_2^a(2\Delta)^2\vert \widehat{\tilde{\boldsymbol S}^a} \vert\widehat{\tilde S_{kk}^a}\delta_{ij}.
\end{equation}

Minimizing the mean square error between the resolved identity and the Smagorinsky model leads to the expression for the coefficient $C_2^a$,
\begin{equation}
  C_2^a=\frac{\langle\mathcal{L}^{*,a}\rangle_C}{\langle\mathcal{M}^{*,a}\mathcal{M}^{*,a}\rangle_C},
\end{equation}
with $\mathcal{M}^{*,a}=-\frac{2}{3}\Delta^2\left[\widehat{\vert\tilde S^a\vert\tilde S_{kk}^a\delta_{ij}}-4\vert\widehat{\tilde S}^a\vert \widehat{\tilde S}_{kk}^a\delta_{ij}\right]$ and operator $\langle\cdot\rangle_C$ representing average over the cell faces.
\section{Averaging procedure}
\label{average}

The given variable $\psi$ can be decomposed into the sum of the Favre-averaged variable $\langle \psi \rangle_F$ and the associated fluctuation $\psi'$. Favre-averaging operations on the variables $\psi^f$ or $\psi^s$ correspond to perform an ensemble average (operator $\langle \cdot \rangle$) of the variables weighted by the ensemble-averaged phase concentration $\langle 1-\phi\rangle$ or $\langle\phi\rangle$ following
\refstepcounter{equation}
$$
  \langle \psi^f \rangle_F = \frac{\langle(1-\phi)\psi^f\rangle}{\langle 1-\bar\phi\rangle}, \quad 
  \langle \psi^s \rangle_F = \frac{\langle\phi\psi^f\rangle}{\langle\phi\rangle}.
  \eqno{(\theequation{\mathit{a},\mathit{b}})}
$$

Numerically, averaged variables are calculated by performing a spatial averaging operation in the streamwise and spanwise directions of a temporally averaged variable $\langle \psi \rangle_t$ following

\begin{equation}
  \langle \psi\rangle = \frac{1}{L_xL_z}\int_0^{L_x}\int_0^{L_z}\langle \psi \rangle_t dx dz,
\end{equation}
with $L_x$ and $L_z$ the lengths of the numerical domain in the streamwise and spanwise directions respectively.

The temporal averaging operation is performed using an iterative procedure at each time step with the temporal average value of the variable $\psi$ at time $t_{n+1}$ given by

\begin{equation}
 \langle\psi(t_{n+1})\rangle_{t} = \frac{\psi(t_{n+1}) + n\langle\psi(t_n)\rangle_t}{n+1}.
\end{equation}

Second-order statistical moments such as r.m.s. of the velocity fluctuations or Reynolds stresses are obtained by calculating the fluid or solid Favre-averaged covariance tensor $\langle\psi'_i\psi'_j\rangle_F$ following

\begin{equation}
 \langle\psi'_i\psi'_j\rangle_F = \langle\psi_i\psi_j\rangle_F - \langle\psi_i\rangle_F\langle\psi_j\rangle_F
\end{equation}
One can notice that in clear water conditions (without solid phase), fluid phase Favre averaging is equivalent to ensemble averaging.

\bibliographystyle{jfm}
\bibliography{jfm}

\end{document}